# PHASE SPACE CORRESPONDENCE BETWEEN CLASSICAL OPTICS AND QUANTUM MECHANICS

D. Dragoman – Univ. Bucharest, Physics Dept., P.O. Box MG-11, 76900 Bucharest, Romania

## ABSTRACT

The paper scrutinizes both the similarities and the differences between the classical optics and quantum mechanical theories in phase space, especially between the Wigner distribution functions (WDF) defined in the respective phase spaces. Classical optics is able to provide an understanding of either the corpuscular or wave aspects of quantum mechanics, reflected in phase space through the classical limit of the quantum WDF or the WDF in classical optics, respectively. However, classical optics, as any classical theory, cannot mimic the wave-particle duality that is at the heart of quantum mechanics. Moreover, it is never enough underlined that, although the mathematical phase space formalisms in classical optics and quantum mechanics are very similar, the main difference between these theories, evidenced in the results of measurements, is as deep as it can get even in phase space. On the other hand, the phase space treatment allows an unexpected similar treatment of interference phenomena, although quantum and classical superpositions of wavefunctions and fields, respectively, have a completely different behavior. This similarity originates in the bilinear character of the WDF in both quantum mechanics and classical optical wave theory. Actually, the phase space treatment of the quantum and classical wave theory is identical from the mathematical point of view, if the Planck's constant is replaced by the wavelength of light. Even WDFs of particular quantum states, such as the Schrödinger cat state, can be mimicked by classical optical means, but not the true quantum character, which resides in the probability significance of the wavefunction, in comparison to the physical 'realness' of classical wave fields.

## 1. INTRODUCTION

Quantum mechanics was born from the need to quantify the energy of the emitted or absorbed electromagnetic radiation in order to explain the black body spectrum and the photoelectric effect. Light was thus considered, from the beginning of the quantum revolution, either as an extended wave or as a bunch of particles with definite energy. The subsequent evolution of quantum mechanics has not succeeded to settle the century old controversy regarding the corpuscular or wave-like nature of light, but rather deepened the mystery, extending the wave-corpuscle controversy to material particles. Quantum mechanics, and more recently quantum optics, is able to correctly calculate the essential characteristics of a quantum system, such as its energy levels, quantum statistics and so on, but still longs for a proper interpretation of its calculations. The widely accepted probabilistic interpretation of quantum mechanics is still an open question, the meaning and even existence of a wavefunction for photons, for example, being still subject to debate.

One of the main reasons for this paradoxical situation of the most successful recent theory in physics is that although it has initially borrowed a lot of concepts from classical physics, in particular from classical optics, the subsequent evolution of quantum mechanics has focused towards developing its own language, manifestly different from classical mechanics. The aim of the present paper is to show how much is classical and how much is quantum in quantum mechanics, and why and where quantum and classical mechanics agree in their predictions. The success of this approach is maximized in the phase space realm, where both classical and quantum theories are expressed in the same mathematical language.

### 1.1. EXISTING ANALOGIES BETWEEN CLASSICAL OPTICS AND QUANTUM MECHANICS

Classical and quantum physics are fundamentally different from a conceptual point of view. For example, the wavefunction is only a probability amplitude in quantum mechanics, whereas in classical optics its analog – the electric field – is a measurable quantity. From the beginning, the founders of the quantum theory tried to find at least a formal connection to classical mechanics. In Schrödinger's view, for example, classical dynamics of point particles should be the "geometrical optics" approximation of a linear wave equation, in the same way as ray optics is a limiting approximation of wave optics (Schrödinger [1926]).

Over the following years some rigorous mathematical analogies between classical optics and quantum mechanics have been identified. One of the better known and widely exploited is based on the similarity between the time-independent Schrödinger equation and the time-independent Helmholtz equation. This analogy led to the design of multilayered optical structures with the same transmission characteristics as their quantum counterparts with 0D, 1D or 2D dimensions (Dragoman and Dragoman [1999]), but with the advantage of a much easier characterization due to their order-of-magnitudes larger dimensions compared to the quantum structures. The same analogy was also employed to show that the transverse modes of aspherical laser resonators are similar to the eigenstates of the stationary Schrödinger equation with a potential well determined by the mirror profile. Although this equivalence holds only for short cavity lengths in comparison with the Rayleigh range of the fundamental mode, higher-order corrections for longer resonator lengths can also be found (Paré, Gagnon and Bélanger [1992]). Nienhuis and Allen [1993] proved that the Hermite-Gauss or the Laguerre-Gauss modes of a laser beam can be described using the operator algebra of the quantum harmonic oscillator. In particular, these modes are generated from the fundamental laser mode by applying ladder operators. In addition, displaced light beams, which are refracted by lenses according to geometrical optics, were found to be the paraxial optics analog of a coherent state. Inhomogeneous graded-index waveguides can also benefit from the quantum-theoretical formalism (Krivoshlykov [1994]). The Franck-Condon principle has found its analogs in paraxial optics in the mismatch of a mode passing through two fibers with different refractive index distributions; an optical analog to the Ramsauer effect has also been identified by Man'ko [1986].

Moreover, the quantum optics squeezed states have been extended to solutions of the Helmholtz equation that contain them in the paraxial approximation (Wolf and Kurmyshev [1993]). The Helmholtz equation reduces in the paraxial approximation to the time-dependent Schrödinger equation, with the distance along the optical axis, the wavelength, and the refraction index playing the roles of time, Planck's constant and potential, respectively, in quantum mechanics. The canonically conjugate momentum describes in the optical case the direction of the ray. Ray optics and classical mechanics are then recovered in the limits $\lambda \to 0$ and $\hbar \to 0$, respectively. Even nonparaxial mappings, corresponding to aberrations in optics, can be considered.

The Fresnel diffraction from a slit has found its equivalent in the two-photon process driven by a chirped pulse (Broers, Noordam and van Linden van den Heuvell [1992]). Theory and experiments show that the analog of the spectral Fresnel zone plate leads, for the case of the two-photon level, to a focusing of spectral energy in a much smaller effective bandwidth than that of the original excitation pulse. This analogy is based on the fact that the diffraction pattern is determined in both cases by the interference between different paths that lead to the same final state. The classical Malus law, which predicts an attenuation of light intensity passing through a linear polarizer with a factor depending on the angle between the polarization direction of the incoming wave and the orientation of polarizer, has also a quantum analog. In the quantum case the same attenuation occurs for spin-1/2 particles detected by a properly oriented Stern-Gerlach apparatus if the statistical averages involve a quasidistribution function that can become negative. This analogy can be extended for arbitrary spin values $s$, the classical limit being obtained for $s \to \infty$ (Wódkiewicz [1995]).

In recent years, classical optics has been completely re-implemented using quantum particles; the newly developed "particle optics" which includes "atom optics" (Meystre [2001]) and "electron optics" (Hawkes and Kasper [1996]) is a testimony of the close relationship between quantum mechanics and classical optics. Incoherent phonons, which propagate ballistically in the crystal, can act as acoustic analogs of classical optical mirrors, lenses, filters or microscopes, generating high-resolution acoustic pictures (see Hu and Nori [1996] and the references therein). It was even demonstrated that some essential properties of quantum information and quantum computation methods are classical wave properties, the quantum nature being unquestionable only in situations where nonlocal entanglement is present (Spreeuw [1998]). The monumental work of Mandel and Wolf [1995] investigates also the relationship between classical and quantum coherence.

The analogies between light propagation and atom optics can be extended beyond the paraxial approximation, to calculate for example consecutive corrections to the "optical Schrödinger equation". This generalized analogy has found applications to the harmonic motion in a graded index fiber and to the tunneling between coupled fibers (Marte and Stenholm [1997]). The relationship between Schrödinger and classical wave propagation was applied to scattering problems. In particular, results valid for electron-impurity scattering were extended to scattering of scalar classical waves from dielectric particles (van Tiggelen and E. Kogan [1994]). In the latter case, the phase-destroying effects, which restrict the observation of interference in multiple electronic scattering to low temperatures or to the mesoscopic regime, are absent.

On the other hand, concepts and phenomena characteristic for propagation of quantum wavefunctions have found their analogs for classical waves. Examples include the optical crystals (Soikoda [2001]) known as photonic band gap structures, the optical Berry phase (Bhandari [1997]), tunneling (Ranfagnie, Mugnai, Fabeni, Pazzi, Naletto and Sozzi [1991]), and even more exotic concepts as weak localization (van Albada and Lagendijk [1985], Akkermans, Wolf and Maynard [1986], Kaveh, Rosenbluh, Edrei and Freund [1986]),

Anderson localization (Anderson [1985], Kaveh [1987]), quantized conductances (Montie, Cosman, 't Hooft, van der Mark and Beenakker [1991]), and conductance fluctuations (Kogan, Baumgartner, Berkovits and Kaveh [1993]).

Last, but not least, functions that have been applied in classical optical problems have been translated in an operator language in quantum mechanics, and vice-versa. One example is the fractional Fourier transform. Although it originates in quantum mechanics (Namias [1980]), it was adopted in classical optics (Lohmann, Mendlovic and Zalevsky [1998]), and then related concepts such as the complex fractional Fourier, developed first in classical optics, have found their way back in quantum mechanics (Chountasis, Vourdas and Bendjaballah [1999]).

As one would expect, despite all these fruitful analogies, there are differences between quantum and classical theories. For example, the long-wavelength classical scattering vanishes as $l^{-4}$, whereas a finite (s-wave) cross-section is obtained in the same limit for Schrödinger potential scattering (van Tiggelen and E. Kogan [1994]). The expectation value of the orbital angular momentum of a paraxial beam of light is expressible not only in terms of an analogous angular momentum of the harmonic oscillator, but also contributions from the ellipticity of the wave fronts and of the light spot are present (Nienhuis and Allen [1993]).

## 1.2. WHY A PHASE SPACE TREATMENT OF THESE ANALOGIES?

Why do we need then a phase space (PS) comparison of quantum mechanics and classical optics? The aim of this work is to show that a PS approach to such a problem can offer a more complete answer to the question of the limit of such analogies. The previously mentioned similarities between quantum and classical phenomena are usually based on the observation of formal resemblance of mathematical equations. What about a small difference – a small, perturbation-like term – in the mathematical formulas? Would this term have any clear interpretation? Probably not. The reason is that classical mechanics and/or classical optics operates with algebraic functions, whereas quantum mechanics works with operators which are applied on different representations of quantum states. There is no obvious connection between these two mathematical theories.

A better understanding of the connection between quantum mechanics and classical optics would be provided if the comparison would be made between similar mathematical languages. Since an operator approach to classical optics, or classical mechanics, would be an un-necessary complication, our attention focuses to a quantum mechanical treatment in the space of numbers. Fortunately, such a treatment has been already developed in the quantum mechanical PS.

## 2. THE PHASE SPACE IN CLASSICAL OPTICS AND QUANTUM MECHANICS
## 2.1 HAMILTONIAN FORMULATION OF THE EQUATIONS OF MOTION
## IN CLASSICAL MECHANICS

The state of a physical system is characterized by the information needed at a given time for the calculation of its future evolution. For a classical particle the state is given by the $n$ independent components of the coordinate vector $\boldsymbol{q} = (q_1, q_2, ... q_n)$ and momentum vector $\boldsymbol{p} = (p_1, p_2, ... p_n)$. They span the $2n$ dimensional PS $(\boldsymbol{q}, \boldsymbol{p})$ of classical mechanics in which the momentum and position vectors are on the same footing and interchangeable. The evolution of a system in PS under the action of a Hamiltonian $H_{cl}(\boldsymbol{q}, \boldsymbol{p})$ is described by a set of first order differential equations for the conjugate variables $\boldsymbol{q}$ and $\boldsymbol{p}$:

$$\dot{q}_i = \partial H_{cl} / \partial p_i, \qquad \dot{p}_i = -\partial H_{cl} / \partial q_i, \tag{2.1}$$

where the dot indicates the total time derivative. For a time-dependent Hamiltonian the above equations are supplemented with $\dot{H}_{cl} = \partial H_{cl} / \partial t$. For systems of classical particles with mass $m$ the Hamiltonian can usually be separated into a kinetic and a potential part $H_{cl} = \boldsymbol{p}^2 / 2m + V(\boldsymbol{q})$, with $V(\boldsymbol{q})$ the potential energy. In this case $p_i = m\dot{q}_i$.

The PS in classical optics depends on the treatment we are using: geometrical (ray) or wave optics. Geometrical optics is an approximation to wave optics which disregards the diffraction, interference or polarization effects, valid whenever the dimensions of various apertures are very large compared to light wavelength. In geometrical optics a Hamiltonian can be defined as $H_{opt} = -[n(\boldsymbol{q})^2 - \boldsymbol{p}^2]^{1/2}$ where $n(\boldsymbol{q}) = n(x, y)$ is the local refractive index and $\boldsymbol{p} = (p_x, p_y) = (nu_x, nu_y)$, with $u_x$, $u_y$ the direction cosines made by a ray with the $x$, $y$ coordinate axes. Equations (2.1) are valid also in this case, with the total time derivative replaced by the derivative with respect to $z$ – the propagation direction of the bunch of rays. Although the Hamiltonian equations are formally similar, the PS of classical mechanics and classical optics are globally different. In classical mechanics the momentum vector $\boldsymbol{p}$ is not restricted in value, whereas in classical optics the form of the Hamiltonian implies that $| \boldsymbol{p} | \leq n$.

Only in the paraxial approximation, when $| \boldsymbol{p} | << n$, $H_{opt} = \boldsymbol{p}^2 / 2n(\boldsymbol{q}) - n(\boldsymbol{q})$ has the same form as the Hamiltonian of classical mechanics. Even in this case, however, classical optics can only be similar to classical mechanics when no light refraction is considered. Otherwise, the discontinuity in the ray direction, similar to an elastic reflection of a particle on a wall, imposes the analogy of geometrical optics to the control theory and not to differentiable classical dynamics (Raszillier and Schempp [1986]). A global map – a third-order Seidel-Lie coma map –exists between $4\boldsymbol{p}$ (or wide-angle) optics, based on the 3D Euclidean algebra, and the paraxial approximation of geometrical optics, based on the Heisenberg-Weyl algebra (Man'ko and Wolf [1989]).

For both classical mechanics and geometrical optics (2.1) can be written as

$$\dot{\hat{\boldsymbol{\iota}}} = \boldsymbol{J} \frac{\partial H_{cl,opt}}{\partial \hat{\boldsymbol{\iota}}}, \tag{2.2}$$

where $\hat{\boldsymbol{\iota}} = \begin{pmatrix} \boldsymbol{q} \\ \boldsymbol{p} \end{pmatrix}$ is the ray vector, and $\boldsymbol{J} = \begin{pmatrix} 0 & I \\ -I & 0 \end{pmatrix}$ is a $2n{\times}2n$ matrix, with $\boldsymbol{I}$ the $n{\times}n$ identity matrix ($n$ = 2 for geometrical optics). This form of the Hamiltonian equations of motion is preserved under canonical transformations, characterized by a matrix $\boldsymbol{M}$, which satisfies the relation

$$\boldsymbol{M}\boldsymbol{J}\boldsymbol{M}^T = \boldsymbol{J}. \tag{2.3}$$

$\boldsymbol{M}$ describes the transformation of variables from the initial set $\hat{\boldsymbol{\iota}}^{\,i}$ to the final set $\hat{\boldsymbol{\iota}}^{\,f}$ and is defined as $M_{ab}(\boldsymbol{x}^i) = \partial \boldsymbol{x}_a^f / \partial \boldsymbol{x}_b^i$. The linear transformations that satisfy (2.3) form the symplectic group. A linear transformation is symplectic in two dimensions if its determinant is 1, additional restrictive conditions being required in higher dimensions (Guillemin and Sternberg [1984]). From a geometrical point of view, Hamiltonian mechanics corresponds to transformations that preserve an antisymmetric, nondegenerate bilinear form defined on an even-dimensional real vector space $\hat{\boldsymbol{\iota}}$, which in two dimensions can be written as $\boldsymbol{s}(\boldsymbol{x}_1, \boldsymbol{x}_2) = q_2 p_1 - q_1 p_2$. In other words, the PS area bounded by a group of trajectories is constant with time; one can directly follow the motion of a bounded region in PS rather than following the individual trajectories that comprised that region.

If instead of one particle we consider now an ensemble of non-interacting particles we can define a probability distribution function in PS $f(\hat{\imath})$. Under the action of the symplectic map the probability distribution function transforms as $f^f(\hat{\imath}) = f^i(M^{-1}\hat{\imath})$. This relation is known as the Liouville's theorem. The Liouville's theorem can be explicitly expressed in terms of the Hamiltonian of the system as

$$\frac{df}{dt} = [f, H_{cl,opt}]_P + \frac{\partial f}{\partial t}, \tag{2.4}$$

where the Poisson bracket is defined by

$$[A, B]_P = \sum_i \left( \frac{\partial A}{\partial q_i} \frac{\partial B}{\partial p_i} - \frac{\partial A}{\partial p_i} \frac{\partial B}{\partial q_i} \right) = \sum_i J_{ij} \frac{\partial A}{\partial z_i} \frac{\partial B}{\partial z_j}. \tag{2.5}$$

The Poisson bracket expresses the symplectic structure of the classical PS. It is a binary antisymmetric relation which, applied on the elements of a commutative ring, makes it into a Lie algebra with the additional requirement that the bracket acts as a derivation of the commutative multiplication. A generalized Liouville theorem exists even for non-Hamiltonian motion, induced by forces that are not derivable from a potential, such as close-range collisions with other species of particles, synchrotron radiation or bremsstrahlung (Lichtenberg [1969]).

(2.4) holds not only for the probability distribution in PS, but for any function $A(q, p)$ of canonical variables. In particular $[q_i, q_j]_P = 0$, $[p_i, p_j]_P = 0$, $[q_i, p_j]_P = d_{ij}$ and (2.1) become $\dot{q} = [q, H_{cl,opt}]_P$, $\dot{p} = [p, H_{cl,opt}]_P$. The expectation value of a PS function for an ensemble of non-interacting particles is defined as

$$\langle A \rangle = \int dq\, dp\, A(p, q) f(p, q). \tag{2.6}$$

In wave optics it is also possible to define a Hamiltonian, but in terms of conjugate functions instead of conjugate vectors. More precisely, expanding the vector potential of the electromagnetic field as $A(q, t) = V^{-1/2} \sum_k \sum_{s=1,2,3} e_{ks} Q_{ks}(t) \exp(ikq)$, where $V$ is the volume in which the field is confined, $k$ is the wavevector, and $e_{ks}$ is a unit vector along the polarization direction $s$, the Hamiltonian of a source-less and current-less electromagnetic field is given by

$H_{em} = (1/2) \sum_{k,s} (P_{ks} P_{ks}^* + w_k^2 Q_{ks} Q_{ks}^*)$ with $P_{ks} = \dot{Q}_{ks}^*$ and $w_k = kc$ (di Bartolo [1991]). The Hamiltonian equations are now

$$\dot{Q}_{ks} = \partial H_{em} / \partial P_{ks}, \dot{P}_{ks} = -\partial H_{em} / \partial Q_{ks}. \tag{2.7}$$

or $\dot{A}(q, t) = [A(q, t), H_{em}]_P$. Since the electric and magnetic fields are related to the vector potential through $E = -\partial A / \partial t$, $B = \nabla \times A$ (the scalar potential of the electromagnetic field is taken as zero), their Poisson brackets are $[E_i(q, t), E_j(q', t)]_P = 0$, $[B_i(q, t), B_j(q', t)]_P = 0$, $[E_i(q, t), B_j(q, t)]_P \neq 0$ (di Bartolo [1991]).

## 2.2. QUANTIZATION PROCEDURES AND THE PHASE SPACE
## OF QUANTUM MECHANICS

The relation between classical and quantum mechanics can be translated in group theory through the relation between the symplectic group and the metaplectic representation, the latter denoting the action of the metaplectic group – double covering of the symplectic group – on the Hilbert space. The quantization of a classical mechanical system reduces then to the association to each quadratic polynomial $H_{cl}$ of a self-adjoint operator $\hat{H}$ acting on the Hilbert space, such that the map from $H_{cl}$ to $-i\hbar^{-1}\hat{H}$ carries the Poisson bracket into commutators (Guillemin and Sternberg [1984]). According to the Groenwald-van Hove theorem, it is not possible, however, to extend the metaplectic representation such as to include nonquadratic polynomials.

More precisely, the quantization of a Hamiltonian nonrelativistic physical system proceeds by raising the classical dynamical variables $\boldsymbol{q}$, $\boldsymbol{p}$ and any function of them $A(\boldsymbol{q}, \boldsymbol{p})$ to the category of linear operators. In doing this, the Poisson brackets $[A, B]_P$ transform to the commutation relations $-i\hbar^{-1}[\hat{A}, \hat{B}] = -i\hbar^{-1}(\hat{A}\hat{B} - \hat{B}\hat{A})$. In particular, in quantum mechanics the position and momentum operators do not commute, i.e. $[\hat{q}, \hat{p}] = i\hbar$. This property of the position and momentum operators, although not encountered in classical mechanics, where the corresponding Poisson bracket vanishes, is not foreign to classical wave optics. In the preceding section we have assigned a non-vanishing Poisson bracket to the electric and magnetic field components of the electromagnetic radiation.

Due to the non-commutativity of $\hat{q}$ and $\hat{p}$, the quantization procedure is unambiguous only when the correspondence between $A(\boldsymbol{q}, \boldsymbol{p})$ and $\hat{A}(\hat{q}, \hat{p})$ is unique, i.e. there is no ambiguity in the ordering rules of the position and momentum operators. Even in this case, however, the canonical quantization privileges the Cartesian frame. This means that, for example, a Hamiltonian expressed in terms of angle-action variables, cannot be quantized in a well-defined manner. Moreover, in the canonical quantization process the Hamiltonian must be identified with the total energy of the system in order to avoid contradictory results. Not even in the $(q, p)$ variables is the Hamiltonian unique for a given motion (Pimpale and Razavy [1988]). However, it was recently shown that PS concepts are essential to define a general procedure of quantization of non-Hamiltonian systems (Bolivar [1998]).

For mixed quantum-classical systems it is possible to define a quantum-classical bracket that reduces to the quantum commutator and the Poisson bracket in the purely quantum and classical cases, respectively. In these systems two distinct sets of variables, with their own Planck constant, correspond to the quantum and classical parts, respectively, so that the Planck constant of the classical part can approach zero leaving the quantum subsystem unchanged (Prezhdo and Kisil [1997]).

In quantum mechanics all information about a quantum state is contained in the state vector, which is a vector in Hilbert space. The observables are Hermitian operators. In the Schrödinger formulation of quantum mechanics the quantum state $|\boldsymbol{\psi}\rangle$ satisfies the linear differential equation

$$i\hbar\frac{\partial |\boldsymbol{\psi}\rangle}{\partial t} = \hat{H}|\boldsymbol{\psi}\rangle \qquad (2.8)$$

with a quantum Hamiltonian $\hat{H}$. The eigenstates $|n\rangle$ of the Hamiltonian are the energy eigenstates. The energy spectrum of a quantum system has usually a discrete and a continuous part. It is usually assumed that a discrete energy spectrum is a manifestation of the quantum nature of a system. But, is this really so? No. Classical optics offers the best counterexample: the propagation constants (energy levels) of classical light in a waveguide have also a discrete as well as a continuous spectrum (Snyder and Love [1983]). Energy discretization appears whenever a constraint is imposed upon the freedom of movement.

Since the eigenstates $|q\rangle$, $|p\rangle$ of the position and momentum operators form complete sets of states, an arbitrary quantum state can be characterized in the position or momentum representation. In the position representation the wavefunction $\psi = \psi(q,t)$ is defined as $\langle q|\psi\rangle$ and the action of the operators $\hat{q}$ and $\hat{p}$ on the quantum state are described by multiplication with $q$ and $-i\hbar\nabla_q$, respectively, so that $\hat{H} = H_{cl}(q, -i\hbar\nabla_q, t)$. In the momentum representation the wavefunction $\psi = \psi(p,t)$ is defined as $\langle p|\psi\rangle$ and $\hat{H} = H_{cl}(i\hbar\nabla_p, p, t)$. Usually, only 1D quantum systems are considered, so that we will restrict from now on to this case. For these systems, $\psi(q)$ and $\psi(p)$ are two representations of the same quantum state; they are related by a Fourier transform $\psi(q) = (2\pi\hbar)^{-1/2} \int_{-\infty}^{\infty} dp\,\psi(p)\exp(iqp/\hbar)$. The squared modulus of the wavefunction in the position or momentum representation gives the corresponding probability density.

We have shown in the previous section that the Hamilton's equations of motion are preserved under linear canonical transformations, which correspond to quadratic Hamiltonians. Quantum mechanical evolution equations are preserved under unitary transformations. The quadratic Hamiltonians in quantum mechanics which generate the rotation around the origin and the squeezing in PS are $\hat{p}^2/2 + w^2\hat{x}^2/2$ and $(\hat{p}\hat{x} + \hat{x}\hat{p})/2$, respectively. The squeezing operator compresses the PS along one coordinate and expands it along the other, transforming one harmonic oscillator into another with different frequency $w$. The other quadratic Hamiltonians, $\hat{p}^2/2$ and $\hat{x}^2/2$ describe in quantum optics the paraxial free propagation of light rays in a homogeneous medium and the action of a thin lens, respectively.

Although the quantum state vector contains all information about the state, it is sometimes difficult, especially in open systems, to use this concept. Therefore, a more appropriate description, which incorporates our lack of knowledge about what pure state the system is actually in, can be given in terms of the Hermitian density operator. It is defined for pure states as $\hat{\rho} \equiv |\psi\rangle\langle\psi|$, or for a superposition of states $|\psi\rangle = \sum_{m=0}^{\infty}\psi_m|m\rangle$ as $\hat{\rho} = \sum_{m,n=0}^{\infty}\psi_m\psi_n^*|m\rangle\langle n| = \sum_{m,n=0}^{\infty}\rho_{mn}|m\rangle\langle n|$. $\rho_{mn}$ are the elements of the density matrix in the energy representation. In a mixed state we cannot describe the state by a superposition, since we only know the probability with which the component states $|m\rangle$ appear. The density matrix is then diagonal in the component-state representation, i.e. $\rho_{mn} = 0$ for $m \neq n$. The density operator has only non-negative eigenvalues (is non-negative) and evolves according to von Neumann equation

$$\frac{d\hat{\rho}}{dt} = -\frac{i}{\hbar}[\hat{H}, \hat{\rho}].$$  (2.9)

$\hat{\rho}$ is extremely useful in expressing the expectation values of an arbitrary operator $\hat{A}$ via the trace operation:

$$\langle\hat{A}\rangle = Tr(\hat{A}\hat{\rho}).$$  (2.10)

For a mixed state (statistical mixture) $Tr\,\hat{\rho}^2 < 1$, while for pure states $Tr\,\hat{\rho}^2 = 1$.

Defining the variance of an operator through $\Delta A = (\langle \hat{A}^2 \rangle - \langle \hat{A} \rangle^2)^{1/2}$, it can be shown that for any pairs of non-commuting operators, in particular for the position and momentum operators

$$\Delta q \Delta p \geq \hbar / 2 \, . \tag{2.11}$$

This uncertainty relation is considered to be the most important difference between classical and quantum mechanics. It says that one cannot simultaneously measure with arbitrary precision the expectation values of two non-commuting operators. In contrast, in classical mechanics the momentum and position of a classical particle can be exactly known at any time. Well, the uncertainty relations are not restricted to the quantum realm. A similar relation (with a different meaning!) exists in wave optics between position and momentum, limited by the light wavelength, while in optical signal processing or Fourier analysis the time and frequency satisfy $\Delta \textbf{\textit{w}} \Delta t \geq 1/2$. This should be expected since, when speaking about position-momentum of photons in free space, position corresponds to time in a retarded frame and the momentum of the photon is related to the frequency via Planck's constant and the velocity of light.

The development and significance of quantum mechanics is checked by the requirement that classical results are recovered in the $\hbar \to 0$ limit. Ehrenfest was the first to show that the equation of motion for the average values of quantum observables coincides with the corresponding classical expression, i.e. $d\langle \hat{q} \rangle / dt = \langle \hat{p} \rangle / m$, $d\langle \hat{p} \rangle / dt = F(\langle \hat{q} \rangle)$. The last equation is valid only if $\langle F(\hat{q}) \rangle \cong F(\langle \hat{q} \rangle)$, with the force defined as $F(q) = -\nabla V(q)$. The validity of Ehrenfest's theorem is neither necessary nor sufficient to identify the classical regime, since the classical limit of a quantum state is not a single classical orbit, but generally an ensemble of orbits. Even when Ehrenfest's theorem fails, a quantum state may behave classically if its evolution is in agreement with the Liouville equation for a regular or chaotic classical ensemble. Thus, a more appropriate criterion for classical behavior is that quantum averages and probability distributions agree, approximately, with the respective classical quantities (Ballentine, Yang and Zibin [1994]). Potentials can, however, be found for which the quantum mechanical motion is identical to the motion of the corresponding classical ensemble. Two classes of such potentials have been founded in Makowski and Konkel [1998].

To difficulty of the quantum-classical correspondence is even more emphasized by the fact that, generally, in order to obtain the correct classical limit when $\hbar \to 0$ the system must be mechanically connected to an infinite number of additional classical degrees of freedom. Quantum effects such as interference or tunneling originate then in the mechanical interactions between different parts of the overall infinite system (Kay [1990]).

Despite all these differences there is a close relation between the classical PS variables and the corresponding quantum operators in the Hilbert space. For example, for any linear transformation in the classical PS described by a symplectic matrix $\textbf{\textit{M}}$ with elements $A$, $B$, $C$, $D$, a unitary operator $\hat{U}(\textbf{\textit{M}})$ can always be constructed such that the same relation exists between the quantum operators: $\hat{q}' = [\hat{U}(\textbf{\textit{M}})]^+ \hat{q} \hat{U}(\textbf{\textit{M}}) = A\hat{q} + B\hat{p}$, $\hat{p}' = [\hat{U}(\textbf{\textit{M}})]^+ \hat{p} \hat{U}(\textbf{\textit{M}}) = C\hat{q} + D\hat{p}$. A special case of such a linear transformation is squeezing (Hong-Yi and VanderLinde [1989]). The evolution of the system changes dramatically, however, in the case of nonlinear evolution. An example of such a situation is the interference of the wavefunction of a trapped atom with Raman-type exciting laser waves, which is the quantum analog of nonlinear optical phenomena such as parametric amplification, multimode mixing and Kerr-type nonlinearities. In this case neighboring PS zones have a different time evolution, the net result being a partitioning of the PS, which may induce strong amplitude squeezing of the motional quantum state as well as quantum interferences (Wallentowitz and Vogel [1997]).

The eigenvalues of the position and momentum operators span the PS of quantum mechanics of massive particles. In quantum optics the electromagnetic field is modeled as a collection of harmonic oscillators with unit mass $m = 1$ and frequency $\boldsymbol{w}$, characterized by bosonic annihilation and creation operators $\hat{a}$ and $\hat{a}^{+}$, respectively. These operators, for which $[\hat{a}, \hat{a}^{+}] = \hbar$, are related to the position and momentum operators of the oscillators as

$$\hat{a} = (2\hbar m \boldsymbol{w})^{-1/2}(m\boldsymbol{w}\hat{q} + i\hat{p}), \quad \hat{a}^{+} = (2\hbar m \boldsymbol{w})^{-1/2}(m\boldsymbol{w}\hat{q} - i\hat{p}). \tag{2.12}$$

However, in quantum optics the position and momentum operators have no clear meanings, so that $\hat{q}$ and $\hat{p}$ are called quadrature operators, and correspond to the in-phase and out-of-phase components of the electric field amplitude. The rotated quadrature operators are linear combinations of the position and momentum operators, with weights determined by the angle of rotation $\boldsymbol{q}$: $\hat{q}_{\boldsymbol{q}} = \hat{q}\cos\boldsymbol{q} + \hat{p}\sin\boldsymbol{q}$, $\hat{p}_{\boldsymbol{q}} = -\hat{q}\sin\boldsymbol{q} + \hat{p}\cos\boldsymbol{q}$.

The quantum mechanical PS, although spanned by variables with the same meaning as in classical mechanics (position and momentum), has a totally different algebraic structure than the classical PS. More precisely, the classical PS, invariant under canonical transformations, is not a metric manifold, since the separation between two points has no invariant meaning. A physically realizable state is characterized in this PS by a density $\boldsymbol{d}(q - q_0)\boldsymbol{d}(p - p_0)$, the motion of this PS point being described by the Hamiltonian equations of motion. The density $\boldsymbol{d}(q - q_0)\boldsymbol{d}(p - p_0)$ is, however, unacceptable in the quantum PS (and in the PS of classical wave optics!), due to the position-momentum uncertainty relation. The quantum PS has thus a metric, non-Riemannian structure, which must coincide with the completely different classical PS in the limit $\hbar \to 0$. Also, the operator algebra of quantum mechanics must be invariant under unitary transformations. To emphasize these differences the quantum PS is called mock PS (sometimes also Weyl PS) and $p$, $q$ are referred to as $c$-numbers. Each rule of association introduces its own $p$, $q$ manifold via a particular choice of the basis set. So, there is a mock PS for each rule of ordering (Balazs and Jennings [1984]).

In the quantum PS it is possible to define a correspondence between a dynamical operator $\hat{A}(\hat{p}, \hat{q})$ and its Weyl image $A(p, q)$. This is essential since there is no isomorphism between the group of canonical transformations in the PS of classical dynamics and the group of unitary transformations in the Hilbert space of quantum mechanics. In the limit $\hbar = 0$, $p$, $q$ become the momenta and coordinates in Cartesian coordinates and the Weyl space becomes the classical PS. Only the Weyl images of linear and quadratic operator functions transform as classical dynamical quantities, the linear inhomogeneous transformations playing a preferred role in the Weyl space endowed with an equiaffine geometry (Balazs [1981]). This conclusion corresponds to that obtained from a group-theoretical point of view.

The group of linear canonical transformations in PS and their applications in various branches of physics are discussed in Kim and Noz [1991]. This group of transformations for $n$ pairs of conjugate variables, characterized by symplectic matrices which can be written as products of translation, rotation and squeezing matrices, is the inhomogeneous symplectic group $ISp(2n)$. Its subgroup, the homogeneous symplectic group $Sp(2n)$ (which does not include translations), is locally isomorphic to the $(2+1)$-dimensional and $(3+2)$-dimensional Lorentz groups for $n = 1$ and $n = 2$, respectively. In quantum optics the $SU(2)$ algebra can be used to calculate the electron-counting probability and the $SU(1,1)$ Lie algebra is useful for the calculation of the photon-counting probability and for the investigation of the quantum-nondemolition measurement of photon number in four-wave-mixing (Ban [1993]).

## 3. DEFINITIONS AND PROPERTIES OF PHASE SPACE
## DISTRIBUTION FUNCTIONS

Since it is not possible to access a $(q, p)$ point in the quantum PS due to the uncertainty principle, we cannot define a localized probability distribution in quantum mechanics, but only quasiprobability distributions that yield correct results for observable quantities. In terms of any distribution function $F(q, p, t)$ the expectation value of an arbitrary operator $\hat{A}(\hat{q}, \hat{p})$ can be calculated similarly as in the classical PS, i.e. as

$$\langle \hat{A}(\hat{q}, \hat{p}) \rangle = Tr[\hat{r}(\hat{q}, \hat{p}, t)\hat{A}(\hat{q}, \hat{p})] = \int dq dp A(q, p) F(q, p, t) \tag{3.1}$$

where $A(q, p)$ is the scalar function obtained by replacing the operators $\hat{q}$, $\hat{p}$ in $\hat{A}$ with scalar variables $q$ and $p$. Different scalar functions $A(q, p)$ and hence different distribution functions are obtained for different rules of ordering of the non-commuting position and momentum operators. All distribution functions contain the same amount of information about the quantum system. The selection of a quasiprobability distribution is merely imposed by the problem to be solved, the principal requirement being usually that of simplicity.

For quantum-classical correspondence problems, the most interesting quasiprobability distribution is the Wigner distribution function (WDF), which corresponds to the Weyl or symmetric rule of association. For pure states the WDF is defined as

$$W(q, p; t) = (2\pi\hbar)^{-1} \int dx \exp(-ipx/\hbar)\psi^*(q - x/2; t)\psi(q + x/2; t) \tag{3.2a}$$

whereas for mixed states

$$W(q, p; t) = (2\pi\hbar)^{-1} \int dx \exp(-ipx/\hbar)\langle q + x/2 \mid \hat{r}(t) \mid q - x/2 \rangle \tag{3.2b}$$

The WDF is limited to $\mid W(q, p; t) \mid \leq 1/\pi\hbar$. In the words of its inventor, the WDF "seems to be the simplest" from all bilinear expressions in the wavefunction which are linear in the expectation values of any sum of a function of coordinates and a function of momenta (Wigner [1932]). There are many books and review papers dedicated to the properties and applications of the quantum WDF, as well as to its relation to other distribution functions. We can only mention a few of them here, such as those written by Moyal [1949], Carruthers and Zachariasen [1983], Hillery, O'Connell, Scully, and Wigner [1984], or more recently the excellent review of Lee [1995a] and the books of Walls and Milburn [1994] and Schleich [2001].

A whole class of $s$-parameterized distribution functions can be obtained from the WDF as (Cahill and Glauber [1968a,b])

$$F(q, p; s) = \exp\left(-\frac{s}{2}\frac{\hbar}{2m\omega}\frac{\partial^2}{\partial q^2}\right)\exp\left(-\frac{s}{2}\frac{m\hbar\omega}{2}\frac{\partial^2}{\partial p^2}\right)W(q, p), \tag{3.3}$$

where $s$ can take also complex values (Wünsche [1996a]). The distribution functions for $s = -1$, 0 and 1 correspond to the antinormal, Wigner and normal distribution functions, respectively ($m = 1$ in quantum optics). For normal ordered operators, the powers of $\hat{a}^+$ precede the powers of $\hat{a}$, the opposite holding for antinormal ordering. The normal distribution function, also called Glauber-Sudarshan or $P$ function, is mostly used in the quantum theory of optical coherence, where expectation values of normally ordered products are of interest.

The distribution function corresponding to $s = -1$, called also $Q$ function, gives the probability distribution for finding the coherent state $|\boldsymbol{a}\rangle$ in the state $\hat{\boldsymbol{r}}$ since $Q(q, p)$ $= \boldsymbol{p}^{-1}\langle\boldsymbol{a}|\hat{\boldsymbol{r}}|\boldsymbol{a}\rangle$. The $Q$ function is always positive and limited to $0 \leq Q(q, p; t) \leq 1/2\boldsymbol{p}\hbar$. Therefore, it is mainly employed in the PS study of chaotic systems, for which the $Q$ function has the smoothest and simplest structure from all distribution functions. In the classical limit of large mean photon numbers the distinctions depending on the ordering of operators vanish and the expressions for $Q$, $P$ become identical.

The $Q$ function is a particular case of a class of non-negative quantum distribution functions – the Husimi functions – obtained by smoothing the WDF with a minimum uncertainty squeezed Gaussian function characterized by a positive constant $\boldsymbol{z}$ :

$$H(q, p, t) = (\boldsymbol{p}\hbar)^{-1} \int dq'\, dp' \exp[-m\boldsymbol{z}(q'-q)^2 / \hbar - (p'-p)^2 / \hbar m\boldsymbol{z}] W(q', p', t) \qquad (3.4)$$

The $Q$ function is retrieved for $\boldsymbol{z} = \boldsymbol{w}$, i.e. when the WDF is smoothed by a coherent state wave packet. The Husimi function is associated with the antinormal ordering of the squeezed photon annihilation and creation operators.

In the coherent state representation, two distributions for different $s$'s are related to one another through a convolution or smoothening operation that depends on the difference in $s$:

$$F(\boldsymbol{a}, s) = \int F(\boldsymbol{b}, s') \left[ \frac{2}{\boldsymbol{p}(s'-s)} \exp\left( -\frac{2\,|\boldsymbol{a}-\boldsymbol{b}|^2}{s'-s} \right) \right] d^2\boldsymbol{b} \qquad (3.5)$$

for $s' > s$. The quasiprobability distributions are in general singular for $s > 0$, i.e. expressed in terms of generalized functions such as delta functions and their derivatives, whereas for $s < -1$ the distribution is well definite and for $s < 0$ is always regular. These behaviors can be understood by viewing (3.5) as an operation of smoothing in the direction of decreasing $s$.

Another class of distribution functions which includes the antistandard, Wigner, and standard distributions for $b = -1$, 0, and 1, respectively, is defined as (see O'Connell and Wang [1985] and the references therein)

$$G(q, p; b) = \exp\left( \frac{i\hbar b}{2} \frac{\partial}{\partial q} \frac{\partial}{\partial p} \right) W(q, p) \qquad (3.6)$$

For standard ordering all powers of $\hat{q}$ precede those of $\hat{p}$, whereas for antistandard ordering all powers of $\hat{p}$ precede those of $\hat{q}$. The distribution function for $b = -1$ is also called Kirckwood or Rihaczek distribution.

Other association rules and corresponding distribution functions are described in Cohen [1966]. Among them are the positive $P$ function, the Rivier or Margenau-Hill ordering for which the distribution function is given by $[G(q, p, -1) + G(q, p, 1)]/2$, the normal-antinormal ordering distribution $[Q(q, p) + P(q, p)]/2$, the Born and Jordan rule of ordering which gives as distribution the product of square modulus of the wavefunction and its Fourier transform. Some of these distributions, as for example, the last one, are not bilinear in the wavefunction.

It is commonly believed that the WDF is the quantum analog of the classical PS probability distribution even for many-body problems (Shlomo [1985]). Why is the WDF a privileged quasiprobability distribution? First, it is a real distribution that can, however, take negative values over certain regions of PS. However, the negative regions of the WDF cannot extend over areas significantly wider than $\hbar / 2$. The realness property is also shared by the $P$ distribution, which is highly nonsingular, and by the $Q$ and Husimi functions, which are in addition non-negative. Then, it satisfies the marginal properties, also satisfied by a classical probability distribution, that

$$\int_{-\infty}^{\infty} dp\, W(q,p) = \langle q \mid \hat{\boldsymbol{r}} \mid q \rangle, \quad \int_{-\infty}^{\infty} dq\, W(q,p) = \langle p \mid \hat{\boldsymbol{r}} \mid p \rangle, \tag{3.7}$$

and the normalization condition $\int_{-\infty}^{\infty} dq\, dp\, W(q,p) = 1$. The correct quantum mechanical marginals are not given, for example, by the $P$, $Q$ and Husimi functions. The marginal distributions of WDF can be directly measured with homodyne or balanced homodyne detection schemes (see next section). For $s$-parameterized distribution functions the marginal distributions are positive for $s \leq 0$, but can take negative values or even be singular for $s > 0$ (Orlowski and Wünsche [1993]).

Another desirable feature of the WDF is that it is invariant with respect to time and space reflections, and it is Galilei invariant, i.e. it transforms as $W(q,p) \to W(q+q',p)$ and $W(q,p) \to W(q,p+p')$ if $\boldsymbol{y}(q) \to \boldsymbol{y}(q+q')$ and $\boldsymbol{y}(q) \to \exp(-ip'q/\hbar)\boldsymbol{y}(q)$, respectively. Then, it is the only quasiprobability distribution that satisfies the overlap property (O'Connell and Wigner [1981])

$$Tr(\hat{\boldsymbol{r}}_1 \hat{\boldsymbol{r}}_2) = 2\boldsymbol{p}\hbar \int_{-\infty}^{\infty} dq \int_{-\infty}^{\infty} dp\, W_1(q,p) W_2(q,p). \tag{3.8}$$

For all other $s$-parameterized quasiprobabilities (3.8) must be replaced by (Leonhardt [1997]) $Tr(\hat{\boldsymbol{r}}_1 \hat{\boldsymbol{r}}_2) = 2\boldsymbol{p}\hbar \int_{-\infty}^{\infty} dq \int_{-\infty}^{\infty} dp\, W_1(q,p;s) W_2(q,p;-s)$. Moreover, WDF was also shown to be the simplest description in nonequilibrium situations. In particular, the WDF has the simplest correspondence to the Bloch equation, which has been extensively used in calculations of quantum corrections to classical distribution functions (O'Connell and Wang [1985]). Muga, Palao and Sala [1998] showed that the WDF is also the closest to the classical probability when average local values and local variances of a quantum observable are numerically compared with their classical counterpart.

The WDF forms also a complete, orthonormal set, in the sense that for a pure function $\boldsymbol{y}(q,t) = \sum a_n(t) \boldsymbol{f}_n(q)$ with $\boldsymbol{f}_n$ the $n$th eigenstate of the system,

$$W(q,p,t) = \sum_{n,m} a_n^*(t) a_m(t) W_{nm}(q,p), \tag{3.9}$$

where $W_{nm}(q,p) = (2\boldsymbol{p}\hbar)^{-1} \int dx \exp(-ipx/\hbar) \boldsymbol{y}_n^*(q-x/2) \boldsymbol{y}_m(q+x/2)$, $\int dq\, dp\, W_{nm}(q,p) W_{n'm'}^*(q,p) = (2\boldsymbol{p}\hbar)^{-1} \boldsymbol{d}_{nn'} \boldsymbol{d}_{mm'}$ and $\sum_{n,m} W_{nm}(q,p) W_{nm}^*(q',p')$ $= (2\boldsymbol{p}\hbar)^{-1} \boldsymbol{d}(q-q') \boldsymbol{d}(p-p')$.

The terms with $n = m$ in (3.9) are called auto-terms, the other being the cross- or interference terms. Note that the WDF for a mixed state is the weighted sum of pure WDFs (it does not contain interference terms); in contrast, the $Q$ function cannot distinguish between statistical mixtures and macroscopic quantum superpositions.

All the properties mentioned above, with the exception of the non-positiveness, are sheared by a classical probability distribution. Therefore, the WDF is called a quasiprobability. This classical-like quantum PS distribution can be identified by using only the postulate that (Bertrand and Bertrand [1987]) $pr(q,\boldsymbol{q}) = \int W(q\cos\boldsymbol{q} - p\sin\boldsymbol{q}, q\sin\boldsymbol{q} + p\cos\boldsymbol{q}) dp$, where $pr(q,\boldsymbol{q})$ is the position probability distribution after an arbitrary phase shift $\boldsymbol{q}$. There is no non-negative distribution, bilinear in the wavefunction, and which yields the correct quantum mechanical marginal distributions (Srinivas and Wolf [1975]). Non-negative Wigner-type distributions for all quantum states can be obtained, however, by smoothing with a Gaussian whose variance is greater than or equal to that of the minimum uncertainty wave packet, or by integrating the WDF over PS regions of the order $\hbar^{3n}$, where $n$ is the number of dimensions (Cartwright [1976]). Non-negative smoothed WDF distributions, which include as a special case the Husimi distributions, can be used to formulate quantum mechanics (Lalovic, Davidovic and Bijedic [1992]).

But is the quantum PS formalism so 'quantum'? Is it related to the PS formalism of classical physics only in the $\hbar \to 0$ limit? Well, no. A WDF formally identical to that defined for pure and mixed quantum states has been long ago defined (and used with considerably success) for coherent (Bastiaans [1979]) and partially coherent classical light beams (Bastiaans [1986]), respectively. The only difference is that the Planck's constant $\hbar$ should be replaced in this case by the normalized wavelength $\mathchar'26\mkern-10mu\lambda = \boldsymbol{1}/2\boldsymbol{p}$ and that the density matrix in (3.2b) should be replaced by the coherence function in classical optics. In rest, all the properties of the quantum and classical WDF defined in this way are identical; for a review of the properties and applications of the WDF in classical optics see Dragoman [1997]. Even the non-positive property of the WDF is preserved in classical optics. Not to mention that signal processing has benefited also from distribution functions defined on the time-frequency PS (Cohen [1989]), in particular from the WDF (Claasen and Meklenbräuker [1980a,b]). The relationship between the quantum and classical WDFs is also supported by the work of Bialynicki-Birula [2000], who showed that for the full electromagnetic field the role of position and momentum is played by the magnetic and electric induction vectors and the analog of the WDF is a functional of $\boldsymbol{B}$ and $\boldsymbol{D}$. Actually, similarities between the quantum and classical WDFs for particular states have been observed by many authors. Serimaa, Javanainen and Varró [1986] even defined a gauge-invariant Wigner operator and a gauge-independent Wigner function that allow for both quantized and classical electromagnetic fields.

Equations (3.2a) and (3.2b) show that the WDF can be calculated from the wavefunction or the density matrix of a quantum system. However, the WDF can be obtained directly in PS by solving a system of coupled linear partial differential equations derived from the time-independent Schrödinger equation. This system is

$$L_1 W(q,p) \equiv \left[ -\frac{p}{m}\frac{\partial}{\partial q} + \sum_{r=1,3,5..}^{\infty} \frac{1}{r!}\left(\frac{i\hbar}{2}\right)^{r-1}\frac{d^r V}{dq^r}\frac{\partial^r}{\partial p^r} \right] W(q,p) = 0 \qquad (3.10a)$$

$$L_2 W(q,p) \equiv \left[ \frac{p^2}{2m} + V(q) - E - \frac{\hbar^2}{8m}\frac{\partial^2}{\partial q^2} + \sum_{r=2,4,6..}^{\infty} \frac{1}{r!}\left(\frac{i\hbar}{2}\right)^r\frac{d^r V}{dq^r}\frac{\partial^r}{\partial p^r} \right] W(q,p) = 0 \qquad (3.10b)$$

where $V(q)$ is the potential energy. Although in general these equations are of infinite order in $p$, only a finite number of derivates contribute for a polynomial potential $V(q)$. The simplest way to extract the WDF from these equations is to expand in into a series of products of Chebyshev polynomials depending on $q$ or $p$ (Hug, Menke and Schleich [1998]).

How can we know that a solution of the equations (3.10a) and (3.10b) is really a WDF, i.e. it corresponds to a certain quantum wavefunction or density matrix? A simple answer is: make sure that the density matrix obtained from the WDF through

$$\boldsymbol{r}(v,u) = \int dp \exp[\, ip(u-v)/\hbar\,] W((u+v)/2, p) \qquad (3.11)$$

have non-negative eigenvalues. More complex criteria exist, however. Narcovich and O'Connell [1986] showed that, besides satisfying the normalization condition, a function $W(q,p)$ should have a continuous and $\hbar$-positive type symplectic Fourier transform, defined as $\tilde{W}(u,v) = \int W(q,p)\exp[i(qv - up)]dq\,dp$, in order to be a WDF. A function $\tilde{W}$ is of $\hbar$-positive type if, for every choice of points $a_1 = (u_1,v_1)$, $a_2 = (u_2,v_2)$, $\ldots a_m = (u_m,v_m)$, the $m \times m$ matrix with elements $\exp[i\hbar s(a_k,a_j)/2]\tilde{W}(a_j - a_k)$ is non-negative. Here $s(a_k,a_j) = u_j v_k - u_k v_j$ is the symplectic form. A quantum state differs from a classical state in that it requires $\tilde{W}(a)$ to be of $\hbar$-positive type instead of positive type in the sense of Bochner. The two conditions of positivity are identical in the limit $\hbar = 0$. The $\hbar$-positive type condition assures that the uncertainty relations are respected, the opposite being not true, i.e. the uncertainty relations alone do not assure that a real PS function is a WDF.

Up to now we have defined the quantum PS as being spanned by the momentum and position coordinates, or by the complex variable $a$ for coherent states. However, it is possible to define a quantum mechanical PS starting from any two mutually incompatible, not necessarily canonically conjugate, complete sets of operators $\hat{A} = \{\hat{A}_1, \hat{A}_2, \ldots\}$, $\hat{B} = \{\hat{B}_1, \hat{B}_2, \ldots\}$ with eigenvalues $a = (a_1, a_2, \ldots)$, $b = (b_1, b_2, \ldots)$. The PS is then spanned by $(a,b) = (a_1, a_2, \ldots, b_1, b_2, \ldots)$ with the variables taking values over the respective continuous or discrete eigenvalue spectra. When it is possible to obtain eigenfunctions of a particular complete set of operators in more than one representation, the relation of one quantum PS to the other is obtained using Dirac's transformation theory. Defining the distribution function associated to a state $|y\rangle$ in the PS corresponding to the complete sets of Hermitian operators $\hat{A}$ and $\hat{B}$ as $F(a,b) = \langle a \,|\, y \rangle \langle y \,|\, b \rangle \langle b \,|\, a \rangle$ and the PS mapping of an operator $\hat{O}$ as $O(a,b) = \langle b \,|\, \hat{O} \,|\, a \rangle / \langle b \,|\, a \rangle$, any transformation to another PS corresponding to the pair of complete sets $\hat{C}$, $\hat{D}$ converts these quantities to

$$f(c,d) = \sum_{a,b} f(a,b)(\langle c \,|\, a \rangle \langle b \,|\, d \rangle \langle d \,|\, c \rangle / \langle b \,|\, a \rangle), \tag{3.12a}$$

$$O(c,d) = \sum_{a,b} O(a,b)(\langle d \,|\, b \rangle \langle a \,|\, c \rangle \langle b \,|\, a \rangle / \langle d \,|\, c \rangle). \tag{3.12b}$$

In particular, the complete sets of operators can be $\hat{q}$ and $\hat{H}$. In this case one finds that the degeneracies of the PS motion are not, in general, reflected in the degeneracies of the energy eigenvalues, and that the PS constants of motion do not always correspond to quantum constants of motion (Pimpale and Razavy [1988]). The WDF has been generalized, in particular, for relativistic spin-zero quantum particles in an external electromagnetic field (Holland, Kyprianidis, Maric and Vigier [1986]), for rotation-angle and angular-momentum variables (Bizarro [1994]), for a general angular-momentum state with applications to collections of two-level atoms (Dowling, Agarwal and Schleich [1994]), and a WDF has even been defined in the number-phase PS with analogous properties to the WDF associated to position and momentum observables (Vaccaro and Pegg [1990], Vaccaro [1995]).

In order to develop in PS a mathematical formalism analogous to that of the Heisenberg equation of motion for a quantum-mechanical operator, and so to deepen the similarities between the PS and Heisenberg treatments of quantum mechanics, operators have also been defined in PS (Ghosh and Dhara [1991]). The Wigner operator $\hat{A}_W(q,p) = A(\hat{Q}, \hat{P})$ is obtained from the WDF function corresponding to an arbitrary quantum mechanical operator $A(q,p) = \int dy \exp(ipy/\hbar)\langle q - y/2 \,|\, \hat{A} \,|\, q + y/2 \rangle$ by replacing $q$, $p$ with the Bopp operators $\hat{Q} = q - (\hbar/2i)(\partial/\partial p)$, $\hat{P} = p + (\hbar/2i)(\partial/\partial q)$. These Wigner operators do not act on the Hilbert space, but on functions in PS; they are not needed for evaluating the expectation values, but can be used to develop time-dependent density-functional theories in PS or master-equations for open quantum systems. A quantum theory using operators can thus be derived in PS, which in the limit $\hbar \to 0$ leads to the canonical formulation of classical mechanics. By introducing operators in the Liouville space, viewed as vectors and represented by kets, on which act superoperators, the PS formulation becomes a representation in a peculiar Liouville-space basis, transforming naturally under Galilean changes of reference frames. In this formalism the rate of change (though not the higher-order time derivatives) of the expectation value of a quadratic operator is the same as if the WDF obeyed a Liouville equation, whatever the Hamiltonian (Royer [1991]). In the bra-ket PS formalism it can be shown that the change of PS representation follows the same rules as a change of representation for a wavefunction, a much simpler transformation rule than in PS approaches based only on functions of PS variables (Wilkie and Brumer [2000]). The role of Galilean space-time symmetries in selecting a certain representation is thus more directly evidenced (Royer [1992]).

## 4. NONCLASSICAL STATES IN PHASE SPACE

The quantum states usually encountered in quantum optics are the coherent states, the Fock states (or number states) and the squeezed states. The properties of all these quantum states are described in detail in any quantum optics textbook; we are concerned here only with their representation in PS. Any of these states can be represented in PS by any of the quasiprobability distributions, in particular by the WDF. In most cases however, they are represented by the so-called 'error-box', or contour lines of the WDF, or, in the case of the number-states, by 'energy-bands'.

The Fock states $|n\rangle$, eigenstates of the photon-number operator $\hat{n} = \hat{a}^+ \hat{a}$, have a definite number of photons, but a completely random phase. The wavefunctions of the energy eigenstates $|n\rangle$ of a harmonic oscillator with mass $m$ and frequency $w$ in the position space are given by $\psi_n(q) = N_n H_n(kq) \exp[-(kq)^2/2]$, where $H_n$ are the Hermite polynomials, $k \equiv \sqrt{mw/\hbar}$ and $N_n = (k^2/\pi)^{1/4}(2^n n!)^{-1/2}$. In the large-$n$ limit (Schleich [2001]) the energy wavefunction takes the form $\psi_n(q) \cong \sqrt{A_n} \exp(if_n) + \sqrt{A_n} \exp(-if_n)$ where the amplitudes are given by $A_n = (k/2\pi)[2(n+1/2) - k^2 q^2]^{-1/2}$ and the phases are $f_n(q) = S_n(q) - \pi/4$, with $S_n(q)$ the PS area enclosed by the vertical line at $kq$ and the circle of radius $\sqrt{2(n+1/2)}$. The amplitudes $A_n$ of the right and the left going waves are equal because the energy wavefunction is a standing wave. Since according to the uncertainty principle a quantum state cannot have a PS area less than $2\pi\hbar$, the energy eigenstate is a band in PS with boundaries determined by the Planck-Bohr-Sommerfeld quantization condition: PS area $2\pi\hbar n$ on its inner boundary and $2\pi\hbar(n+1)$ on its outer. Its PS area is $2\pi\hbar$, as it should be for a pure quantum state (mixed states can occupy larger PS areas). The PS trajectory for the $n$th energy eigenstate, which encompasses the area $2\pi\hbar(n+1/2)$, runs midway through the band. This (Kramer) PS trajectory, described in the dimensionless variables $Q = kq$, $P = p/\hbar k$ by the circle $Q^2 + P^2 = 2(n+1/2)$ corresponds to the classical, harmonic, PS trajectory of a particle with well-defined energy $E_n = \hbar w(n+1/2)$. The collection of Planck-Bohr-Sommerfeld bands for all $n$ values fills out the PS. The PS representation of an energy eigenstate, and the corresponding WDF have been represented in Figs.1a and 2a, respectively. The WDF of the energy eigenstates of the harmonic oscillator $W_n(Q,P) = (-1)^n (\pi\hbar)^{-1} L_n(2(Q^2 + P^2)) \exp(-Q^2 - P^2)$, where $L_n$ are the Laguerre polynomials, depends only on the energy values and therefore is constant along the PS contours of constant energy. In fact, the WDF oscillates in the region enclosed by the classical PS trajectory and exponentially decreases away from it, $W_n(0,0) = (-1)^n/p\hbar$ changing its sign depending on the quantum number being even or odd. For large $n$ numbers the energy-band in PS can be viewed as a contour of the WDF, taken at a sufficiently high value so that the crest and troughs inside the PS trajectory do not appear.

The coherent state $|\mathbf{a}\rangle$ is the complex eigenstate of the annihilation operator $\hat{a}|\mathbf{a}\rangle = \mathbf{a}|\mathbf{a}\rangle$, its amplitude $|\mathbf{a}|$ and phase $\arg \mathbf{a}$ corresponding to the respective quantities of the complex wave amplitude in classical optics. Due to (2.12), the scaled real and complex parts of $\mathbf{a}$ correspond to the PS coordinates $p$, $q$ (the eigenvalues of the position and momentum operators). For a number of different reasons, including the fact that the coherent state expectation value for a single-mode field operator is the same as for the classical electric field of a monochromatic wave, and the fact that the photon distributions for coherent states is Poissonian, coherent states are as close to wave-like states of the electromagnetic oscillator as quantum mechanics allows. This is valid also for statistical mixtures of coherent states (like thermal fields), all other (nonclassical) states being reduced to classical ones by any kind of losses. The coherent state is represented by a minimum-uncertainty Gaussian wavefunction $\mathbf{y}_{coh}(Q) = (\mathbf{k}^2/p)^{1/4}\exp[-2(\mathrm{Im}\,\mathbf{a})^2]\exp[-(Q-\sqrt{2}\mathbf{a})^2/2]$ displaced from the origin by $\sqrt{2}\mathbf{a}$ (displaced vacuum state, with Gaussian quadrature probability distributions with the same width as for the vacuum). Its WDF is $W(Q,P) = (p\hbar)^{-1}\exp[-(Q-\sqrt{2}\,\mathrm{Re}\,\mathbf{a})^2 - (P+\sqrt{2}\,\mathrm{Im}\,\mathbf{a})^2]$, its error box being a displaced (minimum-uncertainty) circular PS area. The coherent states have an indefinite number of photons, and so a more precisely defined phase than number states. They are generated by a highly stabilized laser operating well above threshold. The PS representation, and the WDF of a coherent state are represented in Figs.1b and 2b, respectively.

Analogously, the WDF of the squeezed ground state function $\mathbf{y}_0(Q) = (s\mathbf{k}^2/p)^{1/4}\exp(-sQ^2/2)$ is $W(Q,P) = (p\hbar)^{-1}\exp[-sQ^2 - P^2/s]$, and the momentum and position uncertainties are given by $\Delta q = (\mathbf{k}\sqrt{2s})^{-1}$, $\Delta p = \hbar\mathbf{k}\sqrt{s/2}$. The PS representation is a Gaussian cigar, elongated in one direction and squeezed in the other. Actually not the state, but the fluctuations are squeezed in one variable (momentum or position) at the expense of the other whenever $s \neq 1$. $s = 1$ corresponds to the coherent state. The squeezed quadrature offers the possibility to enhance the quantum measurement limit. Unlike in a coherent state, the quantum fluctuations in squeezed states are no longer independent of phase. Squeezed states are produced in nonlinear optical processes such as the degenerate parametric amplification. A generalized squeezed state $\mathbf{y}_{sq}(q) = (s\mathbf{k}^2/p)^{1/4}\exp[-2s(\mathrm{Im}\,\mathbf{a})^2]\exp[-s(\mathbf{k}x - \sqrt{2}\mathbf{a})^2/2)$ is represented in the PS as a displaced Gaussian cigar (Fig.1c), and its WDF is shown in Fig.2c.

Phase states are also encountered in quantum optics, despite the difficulty of defining a Hermitian phase operator for states of definite phase. The PS representation (Schleich [2001]) of the phase state defined as $|\mathbf{q}\rangle = (2p)^{-1/2}\sum_{m=0}^{\infty}\exp[i(m+1/2)\mathbf{q}]|m\rangle$ is shown in Fig.1d. Fig.1e displays the PS representation of a superposition of coherent states (a Schrödinger cat state) $|\mathbf{y}\rangle = (N/\sqrt{2})(|\mathbf{a}\exp(i\mathbf{q})\rangle + |\mathbf{a}\exp(-i\mathbf{q})\rangle)$, for real $\mathbf{a}$, where $N = \{1 + \cos[\mathbf{a}^2\sin(2\mathbf{q})]\exp(-2\mathbf{a}\sin^2\mathbf{q})\}^{-1/2}$ is a normalization constant.

Are these states real quantum states? Although the Fock or coherent states are defined in the frame of quantum optics, their wavefunctions in the position representation and the corresponding WDFs have the same form as electromagnetic fields in optical waveguides (Dragoman [2000a], Wódkiewicz and Herling [1998]) and fields produced by coherent light sources, respectively (Dragoman and Dragoman [2001], Wódkiewicz and Herling [1998]). This is true also for superpositions of these states. Moreover, rotation and squeezing of the WDF in PS can be realized in classical optics by fractional-Fourier transforming devices (Lohmann, Mendlovic and Zalevsky [1998]) and magnifier systems, respectively. Negative regions exist also for the WDF of the classical optical states, the negative regions being a consequence of the nonlocal character of the fields and not a signature of nonclassical behavior (Dragoman [2000b]). Actually, negative regions for the WDF arise for any (quantum or classical) state that occupies a PS area larger than the minimum allowed value, due to interferences in PS between the neighboring minimum-uncertainty states in which the state can be decomposed (Dragoman [2000c]). The real quantum character is manifested only in the result of measurement.

How can we express in PS the nonclassicality of a quantum state? There is no unique answer to this question. In Buzek and Knight [1995] PS interference between components of the macroscopically distinguishable coherent states lead to nonclassical characteristics. For nonclassical states the WDF takes negative values (this is however theoretically and experimentally infirmed for the WDF in classical optics). Nonclassical states are also defined as those that have a non-positive $P$ distribution. Another definition, and even classification, of nonclassical states is given in terms of the $P$ distribution function, which is related to the density operator in the diagonal representation of coherent states as $\hat{r} = \int d^2 \boldsymbol{a} P(\boldsymbol{a}) |\boldsymbol{a}\rangle\langle\boldsymbol{a}|$. A state is classical if for any Hermitian operator $\hat{A}$, $Tr(\hat{r}\hat{A}) \geq 0$ for every $A(\boldsymbol{a}) \geq 0$, where $\langle\hat{A}\rangle = Tr(\hat{r}\hat{A}) = \int d^2 \boldsymbol{a} P(\boldsymbol{a}) A(\boldsymbol{a})$, and is nonclassical if $Tr(\hat{r}\hat{A}) < 0$ for some $A(\boldsymbol{a}) \geq 0$. The definition of the nonclassical state can even be refined to include weakly and strongly nonclassical states, when the real and imaginary parts of $\boldsymbol{a}$ are specifically taken into account. In particular, for states described by Gaussian WDFs, the onset of squeezing triggers an abrupt change from the classical to the strongly nonclassical regime without passing through weakly nonclassical states (Arvind, Mukunda and Simon [1997]). For the $s$-parameterized distribution functions a criterion of nonclassicity based on negative regions of quasiprobability distributions was derived in Lütkenhaus and Barnett [1995]. They showed that the $s$ parameter associated to a given state has a critical value $s_c$ such that distributions with $s < s_c$ are positive definite, those with $s > s_c$ are indefinite and for $s = s_c$ are positive semidefinite. $s_c$ is thus a measure of the degree of nonclassical behavior ($s_c \geq -1$ since the Q function is always non-negative). The minimum-uncertainty states with Gaussian wavefunctions, which include the coherent and squeezed states, are the most 'classical' since only for them the WDF is positive definite. For them $s_c = 1$. On the contrary, Fock states are nonclassical since their $P$ function contains derivatives of the delta function.

Another criterion defines nonclassical states through the degree of squeezing or sub-Poissonian statistics (antibunching), defined for a single mode field by

$$S = \langle : (\hat{a}\exp(i\boldsymbol{q}) + \hat{a}^+ \exp(-i\boldsymbol{q}))^2 : \rangle - \langle(\hat{a}\exp(i\boldsymbol{q}) + \hat{a}^+ \exp(-i\boldsymbol{q}))\rangle^2 \qquad (4.1)$$

$$Q = (\langle\hat{a}^{+2}\hat{a}^2\rangle - \langle\hat{a}^+\hat{a}\rangle^2)/\langle\hat{a}^+\hat{a}\rangle, \qquad (4.2)$$

respectively, where :: denotes normal ordering. However, neither squeezing nor antibunching provides a necessary condition for nonclasicality. For squeezing $S$ is negative, while for sub-Poissonian statistics $Q$ is negative. Fields with $Q < 0$ have no classical description via the $P$ function, where states with $Q > 0$ (with super-Poissonian statistics) are classical (Mandel [1979]). For the number states of light $Q = -1$. The coherent state is on the borderline between classical and nonclassical states because for it $Q = 0$ (the photon statistics defined by $|\langle n\,|\,\boldsymbol{\psi}_{coh}\rangle|^2$ is Poissonian). When counted, classical particles obey the same statistical law as coherent states, if taken at random from a pool with an average $|a|^2$ each time. A squeezed state can have a photon number distribution broader or narrower than a Poissonian, depending on whether the reduced fluctuations occur in the phase or amplitude component of the field. In particular squeezed vacuum contains only photon pairs (its photon-number statistics vanishes for odd photon numbers), the probability of finding a photon pair being in this case identical to the probability distribution of independently produced particle pairs. By superposing two coherent states, the photon distribution changes from Poissonian to sub-Poissonian, reflecting the nonclassical character of the superposition, revealed also by the negative values of the WDF.

A modified $Q$ parameter can be introduced to characterize nonclassical light even if both $S$ and $Q$ are positive, an example of its utility being the Schrödinger cat states in regions where they exhibit no sub-Poissonian statistics (Agarwal and Tara [1992]). A measure of nonclassicality of a given radiation field can also be defined as the minimum average photon number of the chaotic light that can destroy all the nonclassical properties of the field. Besides sub-Poissonian photon statistics and squeezing, there can be other nonclassical properties; the many they are the smaller is the number of photons necessary to destroy only one of them (Kim [1999]).

A remark should be made on the fact that squeezing of quadrature fluctuations has a classical analog. For example, for a superposition of two coherent classical Gaussian beams, squeezing is due to destructive interference in the $k$-space ($k\ =\ p\,/\,\hbar\,$), which cause the reduction of the $k$-vector bandwidth below the value of the original Gaussian beam (Wódkiewicz and Herling [1998]). In this case $\Delta q$, $\Delta k$ are defined as statistical spreads of the light intensity and its Fourier transform, respectively (or via the respective second-order moments with WDF the weight function) and for a single Gaussian beam have the values $\Delta q = \Delta k = 1/\sqrt{2}$, as for the uncertainty relation for a single coherent state. The degree of squeezing depends on the separation $d$ between the Gaussians:

$$(\Delta q)^2 = \frac{1}{2} + \frac{d^2}{4}\,\frac{1}{1+\exp(-d^2/4)}\,,\ (\Delta k)^2 = \frac{1}{2} - \frac{d^2}{4}\,\frac{\exp(-d^2/4)}{1+\exp(-d^2/4)} \qquad (4.3)$$

An interesting object for studying the transition from classical to nonclassical behavior is the gray-body; as its absorptivity varies from 0 to 1, the gray-body changes from an extremely nonclassical to an extremely classical state (Lee [1995b]).

## 5. MEASUREMENT PROCEDURES OF PHASE SPACE DISTRIBUTION FUNCTIONS IN QUANTUM MECHANICS AND CLASSICAL OPTICS

Since the canonically conjugate position and momentum variables cannot be measured simultaneously in quantum mechanics, is there any chance to measure at least some of the PS distribution functions, or are they only mathematical constructions? Fortunately, there is not only the possibility, but there are at least three methods for measuring PS probability functions (Freyberger, Bardroff, Leichtle, Schrade and Schleich [1997]), as shown schematically in Fig.3.

In the first method, 'slices' through the WDF are obtained by measuring probability distributions of rotated quadratures. This tomographic method is similar to optical tomography, performed on either light beams or light pulses. In the second method the PS is sampled with areas greater than or equal to the value allowed by the uncertainty relation. Simultaneous knowledge of the conjugate variables of the WDF distribution is obtained, although an approximate, smoothed one; the experiments provide actually the $Q$ function. The smoothing is inherently linked to the possibility of performing these approximate measurements. In the third method, called also the ring method, the WDF is obtained measuring its overlap (by photon counting) with different energy eigenstates.

The tomographic method has been applied to the measurement of the WDF for quantum states of either light or matter. The method was proposed by Bertrand and Bertrand [1987] and Vogel and Risken [1989], and is excellently reviewed in the book of Leonhardt [1997]. Although heterodyne tomographic measurements have also been performed, the established method for WDF reconstruction is homodyne detection, especially balanced homodyne detection, which has the advantage of canceling technical noise and the classical instabilities of the reference field. The first practical demonstration of quantum homodyne tomography was performed by Smithey, Beck, Raymer and Faridani in 1993.

In this method the quadratures of the signal beam are rotated to $\hat{q}_{\varphi} = \hat{q}\cos\varphi + \hat{p}\sin\varphi$, $\hat{p}_{\varphi} = -\hat{q}\sin\varphi + \hat{p}\cos\varphi$ by its interference at a balanced 50/50 beam splitter with an intense coherent laser beam, called the local oscillator (LO), which provides the phase reference $\varphi$. The position quadrature, or more exactly $2^{1/2}|\alpha_{LO}|\hat{q}_{\varphi}$, is obtained from the difference between the photocurrents measured by ideal, linear response photodetectors placed in the path of the two beams emerging from the beam splitter. The amplitude of the local oscillator $|\alpha_{LO}|$ is obtained from measurements of the sum of these photocurrents, while $\varphi$ can be varied by adjusting the LO using, for example, a movable mirror. Denoting by $pr(q,\varphi)$ the set of quadrature distributions, the WDF is obtained from the measured rotated quadratures using the inverse Radon transform

$$W(q,p) = -(2\pi^2)^{-1} \mathbb{P} \int_0^{\pi} \int_{-\infty}^{\infty} dx \, d\varphi \, pr(x,\varphi)/(q\cos\varphi + p\sin\varphi - x)^2, \qquad (5.1)$$

where $\mathbb{P}$ denotes the Cauchy's principal value. Any possible losses due to the mismatch between the LO and the quantum light field, due to non-unit detector efficiency, and so on, or any possible amplification of the signal, determine the measurement not of the WDF but of an $s$-parameterized version of it. For example for detectors with a quantum efficiency $\eta$ one measures $\eta^{-1} F(\eta^{-1/2}x, \eta^{-1/2}p, -(1-\eta)/\eta)$ (Leonhardt and Paul [1993]). Quantum tomography measurements have been performed for classical and nonclassical states of the radiation field by Breitenbach and Schiller [1997].

For material particles the mixing mechanism between $q$ and $p$ is provided by the evolution of the system through free space or combinations of lenses and free spaces. Under free evolution, the WDF of the initial state suffers a shear transform in PS, the WDF being reconstructed from a set of marginal distributions (measured with atom detectors) recorded for different evolution times $t_d$. The rotation angle of the quadratures is related to the evolution time by $\varphi = \tan^{-1}(t_d\hbar/mx_0^2)$, where $x_0$ is a scaling length, and the measured marginal distributions are in this case $pr(q,t_d) = pr(q/\cos\varphi, mx_0^2\tan\varphi/\hbar)$. Free evolution gives only access to rotation angles between 0 and $\pi/2$, additional lenses being necessary to access the whole range from $-\pi/2$ to $\pi/2$ (Pfau and Kurtsiefer [1997]).

Several modifications of the 'classical' balanced homodyne detection scheme have been proposed in order to measure the discrete WDF characterizing quantum states of finite-dimensional systems like atoms or spins (Leonhardt [1995]), or for the tomography of a beam of identically prepared charged particles, entering into an electric field which causes harmonic oscillations in transverse direction (Tegmark [1996]). Optical homodyne tomography was also used to measure the number-phase uncertainty relations (Beck, Smithey, Cooper and Raymer [1993]) or the ultrafast (sub-ps) time-resolved photon statistics of arbitrary single-mode weak fields from phase-averaged quadrature amplitude distributions (Munroe, Boggavarapu, Anderson and Raymer [1995]).

Optical tomographic measurements have been long ago used for image reconstruction from projections (Hermann [1980]) of for the reconstruction of the WDF for light beams (Lohmann and Soffer [1994]). Raymer, Beck and McAlister [1994] designed a tomographic method for the determination of the amplitude and phase of either quasimonochromatic scalar electromagnetic or quantum wavefunction of matter waves in a plane normal to propagation direction. The set-up is a classical optical tomography one, formed from two cylindrical lenses, oriented along different directions and at different distances from the input plane. By recording the intensity distribution in the output plane for different focal lengths and distances from the input plane of the two cylindrical lenses, it is possible to reconstruct the field (and the WDF) for coherent electromagnetic fields or pure quantum states, or to reconstruct the two-point correlation function or density matrix for partially coherent electromagnetic fields or mixed quantum states, respectively. Since time and frequency are also non-commuting variables for nonstationary signals, a time-frequency joint probability density cannot be measured directly, but can only be obtained from marginal distributions along rotated directions in the $(t, \boldsymbol{w})$ plane (Man'ko and Vilela Mendes [1999]). The setup of Beck, Raymer, Walmsley and Wong [1993], consists of a succession of dispersive elements and time lenses (Kolner [1994], Godil, Auld and Bloom [1994]), which mix $\boldsymbol{w}$ and $t$.

Quantum tomography can be simplified considerably if the density operator (and hence the WDF) is reconstructed from its normally ordered moments up to the $n$th order. In this case the measurement of quadrature components for only $n+1$ discrete angles are needed (Wünsche [1996b]); we would like to mention that the signal (and the WDF) can be recovered from the beam's moments also in classical optics (Teague [1980]).

In the simultaneous method, $q$ and $p$ are measured at the same time with limited accuracy. The scheme consists of two entangled homodyne apparata (an eight-port homodyne detector), which measure simultaneously the $q$ and respectively $p$ quadratures of two copies of a light beam, obtained using a beam splitter. For this, the LO's of the two homodyne detectors must have a phase difference of $\boldsymbol{p}/2$. The two homodyne detectors can measure simultaneously the $q$ and $p$ variables only when the respective operators $\hat{q} = \hat{q}_s + \hat{A}$ and $\hat{p} = \hat{p}_s + \hat{B}$ commute, i.e. if $[\hat{q}, \hat{p}] = 0$. Since for the signal beam $[\hat{q}_s, \hat{p}_s] = i\hbar$, extra quantum fluctuations represented by the non-commuting operators $\hat{A}$, $\hat{B}$ must be introduced through the vacuum port of the beam splitter. These extra fluctuations, necessary in order to satisfy the uncertainty principle, double the uncertainty product for $\hat{q}$, $\hat{p}$ and produce a fuzzy picture of the measured quadratures. The measured probability distribution is

$$pr(q, p) = \boldsymbol{p}^{-1} \int_{-\infty}^{\infty} \int_{-\infty}^{\infty} W(q', p') \exp[-T(q'-q)^2 / R - R(p'-p)^2 / T] dq' dp' \qquad (5.2)$$

where $R$, $T$ are the reflected, and transmitted probabilities of the beam splitter, respectively. The set-up measures thus the $Q$ function for a balanced beam splitter, for which $R = T$, or the Husimi function for an unbalanced one. $T$ controls the squeezing of the resolution of the signal WDF. As for optical tomography, a more detailed description reveals that an $s$-parameterized distribution is measured when detection losses appear. For non-unit detection efficiency $\boldsymbol{h}$, $s = -(2 - \boldsymbol{h}) / \boldsymbol{h}$ and the minimum uncertainty product becomes $(1 - s)\hbar / 2$ (Leonhardt, Böhmer and Paul [1995]). The eight-port homodyne detection scheme works even in the case when the input mode contains only one photon (Jacobs and Knight [1996]).

The $Q$ function of multi-particle states, such as Bose-Einstein condensates, can be obtained from the measured atom count probabilities at the output of an atomic interferometer based on Raman transitions between two hyperfine states, if both the phase and transmission parameters of the interferometer are varied (Bolda, Tan and Walls [1998]). The $Q$ function for a light signal can also be determined from the probability of there being no photons in the amplified signal field. This method is based on the fact that in PS linear amplification is a convolution of the signal to the idler field; in this case the idler field is coherent (Kim [1997]). Cascaded optical homodyning was also proposed as a method to determine the PS distributions of optical fields in which the output photon-number statistics of an unbalanced homodyne detection scheme is measured by phase-randomized balanced homodyning (Kis, Kiss, Janszky, Adam, Wallentowitz and Vogel [1999]). The complex amplitude of the local oscillator controls in this case the PS point of interest and a sampling function maps the measured quadrature statistics onto a PS distribution.

Richter [2000] has shown that a slight modification of the eight-port homodyne detection scheme allows the direct measurement of the WDF. The modified set-up consists of a 50/50 beam splitter that splits the signal, followed by a photon counter at one output beam (output mode 1), while the other output beam (output mode 2) forms the input of an eight-port balanced homodyne detection, which measures the $Q$ function of the state. If $P(q,p,n)$ is the joint probability to measure $n$ photons in output mode 1 and the quadrature components $q$ and $p$ in output mode 2, the WDF of the input field is given by $W(q,p)$ $= \sum_{n=0}^{\infty} (-2)^n P(2q,2p,n) \exp[3(q^2+p^2)]$ in normalized coordinates for which $\hbar = 1$.

A classical equivalent of the quantum simultaneous measurement method does not really exist, since in classical optics, although the fields occupy an area in PS greater than a minimum value determined by the wavelength of light, the WDF can be measured in principle with arbitrary high resolution. However, to mimic the quantum measurement, Wódkiewicz and Herling [1998] proposed a method to determine the $Q$ and Husimi functions in classical wave optics by simply smoothing the optical WDF by masks with Gaussian transmittance.

The simultaneous method tells us that in quantum PS the results can always be expressed as a convolution of the WDF of the quantum state with a distribution of the possible states of the measurement device, always distributed over a PS area of at least the order of $\hbar$; the overlaps of two WDFs is always a positive quantity, as follows from the overlap principle of WDF (in contrast, negative regions of the WDF have been directly measured in classical optics; see below). No structures of the WDF finer than $\hbar$ can be directly observed in real measurements, even if these exist (Zurek [2001]). The measurement, or filtering device is needed to resolve the current position and momentum of the investigated system (Wódkiewicz [1984]).

The algorithm for WDF recovery is the simplest in the ring method. It is based on a method proposed by Royer [1977], who showed that the WDF at a point $(q_0, p_0)$ is the expectation of the displaced parity operator $\hat{\Pi}$: $W(q_0,p_0)$ $= \boldsymbol{p}^{-1} \langle \boldsymbol{y} | \hat{D}(q_0,p_0) \hat{\Pi} \hat{D}^+(q_0,p_0) | \boldsymbol{y} \rangle$, where $\hat{D}(q,p) = \exp(ip\hat{q} - iq\hat{p})$ is the coherent displacement operator which displaces a state across PS by an amount $(q,p)$. The motional state of a trapped ion, for example, can be coherently displaced by applying an oscillating (resonant) field that couples to the ion's motion. The amplitude and the phase of the applied field with respect to that of the initial motional state of the ion determine the point $(q_0, p_0)$. For a harmonic oscillator the energy eigenstates are also parity eigenstates, so that the WDF can be obtained from the measured probability distribution of energy eigenstates $P_n(q_0, p_0)$ as $W(q_0,p_0) = \boldsymbol{p}^{-1} \sum_{n=0}^{\infty} (-1)^n P_n(q_0,p_0)$. The measurement of $P_n$ is performed indirectly, coupling the external motion to internal hyperfine levels of the ion via a resonant laser radiation applied for a duration $\boldsymbol{t}$, and by actually measuring the population of one of the hyperfine levels after $t$. Measurements have been performed for different motional states of the ion, such as different harmonic oscillator states, thermal, coherent, squeezed and energy eigenstates, and even superposition of them including Schrödinger cat states; these states are prepared by applying laser pulses and RF fields to a ion in the ground state of the harmonic oscillator (Leibfried, Meekhof, King, Monroe, Itano and Wineland [1996], Leibfried, Pfau and Monroe [1998]). Fields inside high-Q microwave cavities benefited from the same method of measuring their WDF. These fields can be displaced in PS by driving the cavity with a strong coherent field, and can be then probed by a two-level atom prepared initially in one of its energy levels. The measured quantity is the atomic inversion after resonant atom-field interaction for specific interaction times (Kim, Antesberger, Bodendorf and Walther [1998], Lutterbach and Davidovich [1997]). The same method was applied for determining the WDF of molecular vibrational states from measurements of fluorescence after applying a sequence of electromagnetic pulses to a molecule (Davidovich, Orszag and Zagury [1998]). A simpler experimental scheme would imply the exploitation of selection rules for Raman transitions following the coupling of the motional state with the internal levels by Raman laser pulses (Bardroff, Fontenelle and Stenholm [1999]). An endoscopic tomography of single modes of the radiation fields in the cavity, without taking them out, can be performed coupling it with a quantum-nondemolition Hamiltonian to a meter field in a highly squeezed state. Information about the signal field can be obtained from balanced homodyne tomography on the meter field only in out-of-phase measurements, during which the initial state is however changed (Fortunato, Tombesi and Schleich [1999]). Banaszek, Radzewicz, Wódkiewicz and Krasinski [1999] proposed a scheme to implement the ring method for light fields. The coherent displacement is created in this case by a high-transmission beam splitter, which mixes the signal beam with a probe beam, whose amplitude and phase (and thus the point in PS) can be controlled by amplitude and phase electro-optic modulators, respectively. Ideally, the WDF should be obtained as $W(\boldsymbol{b}) = (2/\boldsymbol{p})\sum_{n=0}^{\infty}(-1)^n P_n(\boldsymbol{b})$, where $P_n(\boldsymbol{b})$ is the photon counting statistics of the signal transmitted by the mixing beam splitter for a PS point $\boldsymbol{b}$. However, the $s$-parameterized quasidistribution function with $s = -(1 - \boldsymbol{h}T)/\boldsymbol{h}T$ is obtained instead due to non-unit quantum efficiency $\boldsymbol{h}$ of the detectors. $T$ is the transmission of the mixing beam splitter. If the signal and the probe fields do not match perfectly at the beam splitter, a scaling of the PS point occurs and $\sum_{n=0}^{\infty}(-1)^n P_n(\boldsymbol{b})$
$= (\boldsymbol{p}/2\boldsymbol{h}T)\exp[-2(1-\boldsymbol{x})|\boldsymbol{b}|^2]W(\sqrt{\boldsymbol{x}/\boldsymbol{h}T}\,\boldsymbol{b}, -(1-\boldsymbol{h}T)/\boldsymbol{h}T)$, where $\boldsymbol{x} = V/(2-V)$ is the squared overlap of the two modes expressed in terms of the fringe visibility $V$.

The principle of superposing coherently displaced replica of the signal in order to obtain the WDF is also used in classical optics. The measurement itself, however, does not rely on photon counting, so it is not directly linked with the ring method; what is measured in this case is directly the light intensity after passing through a setup that Fourier transforms the superposition of the two displaced replica of the signal. Measurements of the classical WDF for one-dimensional and two-dimensional, coherent or incoherent light beams have been performed (Bamler and Glünder [1983], Bartelt, Brenner and Lohmann [1980], Brenner and Lohmann [1982], Conner and Li [1985], Iwai, Gupta and Asakura [1986], Weber [1992]), and even a set-up has been proposed to directly measure the WDF of light pulses (Dragoman and Dragoman [1996]). Since direct measurements of WDF in classical optics have revealed regions of negative values (see for example Brenner and Lohmann [1982]), it is sensible to conclude that their presence arise merely due to the nonlocal character of either quantum wavefunctions or classical fields. This conclusion has received an unexpected backing by the measurement of the WDF of a superposition of two coherent Gaussian beams, which has the same form as the WDF of a superposition of two quantum coherent states (a Schrödinger cat state). In dimensionless coordinates $Q$, $P$ the WDF of a superposition of displaced Gaussians $\boldsymbol{y}(Q) \approx \exp[-(Q-d)^2/2] + \exp[-(Q+d)^2/2]$ is given in both quantum and classical cases by

$$W(Q,P) = W_0 \exp(-P^2)\{\exp[-(Q-d)^2] + \exp[-(Q+d)^2] + 2\exp(-Q^2)\cos(2Pd)\} \qquad (5.3)$$

where $W_0$ is a normalization constant. Negative regions arise from the last, interference term. The measured WDF (its modulus) is shown in Fig.4. The interference term in the middle is only present for coherent classical light; for incoherent sources the same superposition (the fields are no longer Gaussian) looks like in Fig.5. The middle term has no longer an oscillatory behavior, but is a simple incoherent superposition of the outer, individual source terms. It should be noted that superpositions of incoherent classical sources have not the same PS representations as statistical mixtures of quantum states. In the latter case no middle interference term should exist at all (Dragoman and Dragoman [2001]). The total or partial disappearance of the interference term due to decoherence (Giulini, Joos, Kiefer, Kupsch, Stamatescu and Zeh [1996]), can be mimicked in classical optics by filtering away the middle term of the WDF.

Experimental methods have been heavily employed not only for the measurement of PS distribution functions, but also for distinguishing coherent superposition of states from statistical mixtures. These can be inferred from the form of the WDF in some cases, but an unambiguous detection method was shown to exist based on the observation of time-dependent spectrum of spontaneous emission from the studied system. The application for the case of a diatomic molecule is discussed in more details in Walmsley and Raymer [1995].

We have insisted here only on the measurement of the WDF and $Q$ function. Due to its highly singular character the $P$ function cannot be generally determined experimentally, but other PS distribution functions have been determined as well. An example is the positive $P$ distribution, measured in a four-port arrangement (Agarwal and Chaturvedi [1994]).

## 6. PROPAGATION OF CLASSICAL FIELDS AND QUANTUM STATES IN PHASE SPACE

The evolution law for the quantum WDF follows directly from the Schrödinger equation satisfied by the wavefunction, or from the von Neumann equation satisfied by the density operator. In either case, for a classical Hamiltonian $H(q,p) = p^2/2m + V(q)$ one obtains (Lee [1995a]):

$$\frac{\partial W}{\partial t} + \frac{p}{m}\frac{\partial W}{\partial q} - \frac{\partial V}{\partial q}\frac{\partial W}{\partial p} = \sum_{n=1}^{\infty} \frac{(\hbar/2i)^{2n}}{(2n+1)!} \frac{\partial^{2n+1} V}{\partial q^{2n+1}} \frac{\partial^{2n+1} W}{\partial p^{2n+1}}, \qquad (6.1)$$

the left-hand side of which is identical to the classical Liouville equation. The 'nonclassical' terms in (6.1) (the right-hand side) contain higher derivatives of the WDF, implying that the PS motion is no longer described by substitution of one PS point for another, as in the canonical transformation, but is analogous to a diffusion process in which a localized function spreads out. It is not a true irreversible diffusion process, which would be described by even derivatives of the WDF, but a reversible quasidiffusion involving only odd derivatives of the WDF (Bohm and Hiley [1981]).

When not all derivatives in (6.1) exist an alternative, integral form can be used:

$$\frac{\partial W}{\partial t} = -\frac{p}{m}\frac{\partial W}{\partial q} + \int_{-\infty}^{\infty} dj W(q, p+j; t) J(q, j), \qquad (6.2)$$

with $J(q,j)=(i/2\pi\hbar^2)\int_{-\infty}^{\infty}dy[V(q+y/2)-V(q-y/2)]\exp(-iyj/\hbar)$. Equation (6.2) reveals again that the PS motion involves an integral over a set of discrete jumps (nonlocal transformations) in momentum, a concept completely different from the continuity of movement implied by classical mechanics (Wigner [1932]).

The evolution law for the WDF distinguishes it again from the other PS distribution functions, which all have an additional term (or terms) even for the free particle case, and so a nonclassical evolution. These $\hbar$ dependent additional terms are not even of quantum origin (O'Connell, Wang and Williams [1984]). The classical-quantum correspondence with respect to PS dynamics seems again to be best studied in terms of the WDF.

The classical limit of (6.1) cannot be obtained simply putting $\hbar=0$, because quantum contributions (and even a $\hbar$ dependence) may be hidden in the expression of the WDF at the initial time $t_0$, $W_0(q,p)$. For a harmonic potential all -dependent terms in (6.1) disappear and the quantum WDF satisfies an equation of motion identical to a classical probability distribution. The quantum effects are hidden in the initial conditions. Note that all quantum corrections (the terms in the right-hand side of (6.1)) depend only on even powers of .

To separate the two different origins of dependence in WDF, the -dependent terms in (6.1) can be replaced by $a\hbar$, where $a$ is a dimensionless parameter, so that classical evolution for WDF (the causal approximation) implies $a=0$, and the classical limit of the WDF is afterwards obtained by additionally taking the $\hbar\to0$ limit. The first-order quantum correction to the causal approximation, called also the quasi-causal approximation (Bund, Mizrahi and Tijero [1996]), implies the consideration of the first-term on the right-hand-side of (6.1) (second order in $a\hbar$). The first-order quantum correction for the PS Thomas-Fermi and Bose distribution functions for fermions and bosons, respectively, has been calculated in Smerzi [1995]. Both distribution functions reduce in the high temperature limit to the Gibbs-Boltzmann distribution.

In the causal approximation each PS point of the initial WDF evolves classically (although reversed in time) as $W_{cl}(q,p;t)=W_0(q(t_0-t),p(t_0-t))$, following a trajectory according to the classical Hamiltonian equations. $W_{cl}$ is also called semiclassical distribution function. If the classical limit of the WDF exists, one obtains the classical distribution function $\delta(q-Q(t))\delta(p-P(t))$, with $Q(t)$, $P(t)$ solutions of the Hamilton equations. It is important not to confound the classical limit of the quantum WDF with the WDF in classical optics!

$W_{cl}$ depends only on the energy associated with the classical trajectory. For a bound state in a potential that vanishes at infinity $W_{cl}(p,q)$ vanishes for points of positive energy, and so corresponds to a stationary distribution of particles trapped inside the potential. Studying different types of potentials, it was found that the $W_{cl}$ curves tend to move away from the regions where the quantum WDF has negative values. In particular, the region where WDF is negative is located in the domain where $W_{cl}$ vanishes (Bund and Tijero [2000]).

The above results concerning the propagation law of the WDF can be reformulated in the following way: when a symplectic matrix is associated to the canonical transformation that sends the initial set of points $x=(q,p)$ in the PS to another final set of points, the corresponding quantum WDF transforms as $W^f(x)=W^i(M^{-1}x)$ (as the classical probability distribution!) only when $M$ is a linear transform (Littlejohn [1986]). Since, according to its definition, $\det M=1$, the area of localization of the WDF is invariant under linear transformations, although it may change its shape. This area of localization cannot be smaller than Planck's constant in quantum mechanics (Kim and Noz [1991]), and cannot be smaller than the wavelength for the WDF in wave optics. The symplectic group Sp(2n,R) technique was used to study the evolution of pure Gaussian quantum states of systems with $n$ degrees of freedom under quadratic Hamiltonians (Simon, Sudarshan and Mukunda [1988]). This evolution was shown to be compactly described by a matrix generalization of the Möbius transformation.

For nonlinear symplectic maps, such as, for example, aberrations in (ray) optics and particle dynamics (Lichtenberg [1969]), the WDF does not follow the simple relation $W(\boldsymbol{x}) = W(\boldsymbol{M}^{-1}\boldsymbol{x})$. In particular, the WDF is not exactly conserved along straight trajectories even in free space, unless the paraxial limit is considered (Wolf, Alonso and Forbes [1999]). The effect of third order aberrations to the paraxial approximation, described by the Hamiltonians $H = p^4$ for the spherical aberration, $H = p^3 x$ for coma, $H = p^2 x^2$, $H = px^3$ and $H = x^4$ for astigmatism, distortion, and pocus, respectively, is described in Rivera, Atakishiyev, Chumakov and Wolf [1997]. For all these cases numerical simulations showed that the classical and the quantum Gaussian WDF evolve differently, and quantum oscillations (including negative WDF regions) appear at the concavities of an initially Gaussian WDF (the Gaussian shape is not preserved in nonlinear evolution). They are caused by self-interference in PS between different parts of the WDF, and have a smaller area than that of the vacuum state. Only the 'top' of the Gaussian Wigner function moves in agreement with classical dynamics. In both classical and quantum mechanics the uncertainty relations are not conserved at nonlinear transformations; these relations are thus a measure of nonlinearity, and not of nonclassicality, and cannot describe an element of PS volume. Such a PS volume measure can be provided by the moments $I_k(t) = (k/2^{k-1})\int W^k(p,x;t)dpdx/2\boldsymbol{p}$ of the WDF, which are constant under classical canonical transformations, but are only preserved in the quantum case under linear transformations. Self-interference is also observed at nonlinear propagation through a Kerr medium described by $H = (1/2)(p^2 + \boldsymbol{w}^2 x^2) + (\boldsymbol{c}/\boldsymbol{w}^2)(p^2 + \boldsymbol{w}^2 x^2)^2$. In this case, standing waves along a circle in PS form at certain times $\boldsymbol{a} = L\boldsymbol{p}/M$, where $L$, $M$ are mutually prime integers. The interference pattern in PS at these moments, known as the $M$-component Schrödinger cat, is associated with pronounced peaks of the WDF moments (Rivera, Atakishiyev, Chumakov and Wolf [1997]).

Note that in the quantum PS spanned by the annihilation and creation operator eigenvalues, the matrix relation satisfied by the WDF has another form. For example, an optical device which transforms linearly an input multimode field with annihilation and creation operators $\hat{a}_i$, $\hat{a}_i^+$ into a field with annihilation and creation operators $\hat{b}_j$, $\hat{b}_j^+$ can be described by $\hat{b}_a = \sum_b (M_{ab}\hat{a}_b + L_{ab}\hat{a}_b^+)$, where the matrices $M$, $L$ satisfy the relations $M\tilde{L} - L\tilde{M} = 0$, $MM^+ - LL^+ = 1$. For this linear transformation, an example of which is a parametric amplifier, the WDF of the output field is given by

$$W_0(z) = W_i([M - L(M^*)^{-1}L^*]^{-1}z - [M^*L^{-1}M - L^*]^{-1}z^*), \qquad (6.3)$$

where $W(z) = \boldsymbol{p}^{-2}Tr[\hat{\boldsymbol{r}}\int d^2\boldsymbol{a}\exp\{-[\boldsymbol{a}(z^* - \hat{a}^+) - \boldsymbol{a}^*(z - \hat{a})]\}]$. The output WDF is thus dependent on the phase of $z$, even if the input WDF is phase-independent (Agarwal [1987]).

Despite all these analogies, in most circumstances the dynamical time evolution of the WDF does not reduce to classical dynamics even if $\hbar \to 0$. Especially for highly coherent density matrices a direct $\hbar$-expansion treatment of quantum corrections is generally not possible, unless a selective re-summation of the terms in the series for the quantum PS propagation is made, case in which a revised or renormalized classical-like dynamics is obtained (Heller [1976]).

The relation between classical and quantum dynamical theories for the WDF is especially interesting when chaotic systems are considered. Then, the degree of non-integrability of the system plays an important role in the quantum-classical comparison. The PS treatment of chaotic systems cannot be briefly summarized here. Therefore, we refer the readers to the review papers of Eckhardt [1988], Bohigas, Tomsovic and Ullmo [1993] and Wilkie and Brumer [1997a,b] for more details. We only point out here that the PS dynamics in the quantum case is determined by the PS dynamics of the classical system. In general the quantum eigenstates can be separated in regular and irregular groups. In integrable systems the wavefunctions peak on the invariant tori quantized by the discrete values of the action, in the regular (quasi-integrable) case chaotic trajectories alternate densely with regular trajectories, exploring each a tiny fraction of the energy space, behavior which is generally valid also for the regular portions of PS for systems with soft chaos, where there is a mixture of integrable and chaotic motions. However, no simple relation exists between classical and quantum PS motion for strongly chaotic systems. Breakdown of the quantum-classical correspondence in chaotic systems is predicted in some papers (Ford and Mantica [1992]), while in others it is argued that even in chaotic systems semiclassical methods are successful for quite long times (Tomsovic and Heller [1991], Provost and Brumer [1995]). In general, due to the uncertainty principle, the quantum WDF cannot resolve the details of the classical trajectories for long evolution times (Berry and Balazs [1979]), so that, depending on the stable or unstable character of the classical motion, the quantum PS dynamics becomes distinct from the classical PS dynamics in longer or shorter times. The rapid divergence between quantum and classical dynamics in integrable systems for initial conditions near an unstable point is caused by the coherent interference of fragments of the wave packet occurring on a short time scale, of the order of the system dynamical time. For a periodically kicked 1D particle, for example, the differences between classical and quantum motions become discernible on a time scale of the order of $\hbar^{-2/3}$, the discrete quasienergy spectrum, and hence a structure in the quantum motion, being identifiable on longer time scales of about $\hbar^{-1}$ (Jensen [1992]).

In some cases the behavior of quantum and classical chaotic systems is different. Quantum systems behave totally different than classical systems, for which the PS is divided by separatrices and stochastic webs into regions in which different types of motion are exhibited (Torres-Vega, Møller and Zúñiga-Segundo [1998]). Moreover, unlike classical stationary distributions, quantum eigenstates can become localized due to the slowness in the rate with which the nonstationary PS distribution sweeps out the available PS (Heller [1987]).

Ballentine and McRae [1998] showed that for chaotic and regular motions of the Hénon-Heiles model the centroid of the quantum state approximately follows the classical trajectory for very narrow probability distributions, the difference between the equations of motion for quantum and classical PS moments scaling as $\hbar^2$. The differences between the quantum and classical dynamics grow exponentially for chaotic motion and as $t^3$ for regular motions, the corresponding difference between the variances of the PS distributions growing also exponentially but with a larger exponent, and as $t^2$, respectively. Quantum interference patterns in PS can be induced by the quantum oscillations of internal degrees of freedom in multicomponent systems, the chaotic dynamics destroying the coherence of the quantum oscillations (Tanaka [1998])

Habib, Shizume, and Zurek [1998] showed that for initial states chosen as Gaussian packets that randomly sample the chaotic part of the PS, a smooth quantum-to-classical transition in nonlinear dynamical systems occurs via decoherence. Decoherence (Giulini, Joos, Kiefer, Kupsch, Stamatescu and Zeh [1996]) destroys the quantum interference with a degree determined by the interplay between the dynamics of the system and the nature and strength of the coupling with environment, and washes out the fine structure in the classical distribution.

So, the quantum and classical predictions, in particular the expectation values of the corresponding variables, become identical.

### 7. INTERACTIONS OF CLASSICAL FIELDS AND QUANTUM STATES AS PHASE SPACE OVERLAP

The most important application of PS overlap is to calculate the transition probability between two quantum states, characterized by density matrices $\hat{r}_1$ and $\hat{r}_2$. According to the overlap principle the transition probability can be written in terms of an overlap between the corresponding WDFs as

$$P_{12} = Tr(\hat{r}_1 \hat{r}_2) = 2p\hbar \int_{-\infty}^{\infty} \int_{-\infty}^{\infty} dq dp W_1(q,p) W_2(q,p) \tag{7.1}$$

A similar formula holds also in classical statistics if $W_1$ and $W_2$ are the PS densities for the classical states before and after the transition. In classical optics (7.1) is used for defining the coupling coefficient between coherent light sources and optical waveguides, a slightly modified relation holding for the case of partially coherent light sources (see Dragoman [1997]). PS overlaps are also encountered in PS matching problems for particles passing through several set-ups (Lichtenberg [1969]).

The quantum transition probability can be measured by the balanced homodyne technique discussed in the previous section, as (Leonhardt [1997])

$$P_{12} = -p^{-1} P \int_0^p \int_{-\infty}^{\infty} \int_{-\infty}^{\infty} d\boldsymbol{q} dq_1 dq_2 \, p_1(q_1,\boldsymbol{q}) \, p_2(q_2,\boldsymbol{q}) / (q_1 - q_2)^2 \tag{7.2}$$

Alternatively, the direct measurement of overlaps of two WDFs can be achieved for two single-mode light beams via photon counting, the PS overlap of the signal and probe beam being realized by a beam splitter (Banaszek and Wódkiewicz [1996]).

The calculation of transition probabilities using the overlap of the WDFs has been applied, for example, in the study of Franck-Condon transitions (Schleich, Walther and Wheeler [1988], Dowling, Schleich and Wheeler [1991]), for defining transition probabilities of momentum jump after a finite-time evolution of the quantum system (Takabayashi [1954]), or, generally, for studying the transition probabilities of quantum-mechanical oscillators for either large or small perturbations (Bartlett and Moyal [1949]). The transition probabilities for coherent and squeezed states from their WDFs are given in Han, Kim and Noz [1989]. Kim and Wigner [1989] showed that the correct form-factor behavior in the harmonic-oscillator model for hadrons can also be traced to the overlap of two Lorentz-deformed PS distribution functions. Recently, a PS formulation of the Fermi's golden rule have been given in Dragoman [2000d], which showed that even for time-dependent interactions described by a Hamiltonian $H_{\text{int}}(q,p)$ the transition probability between an initial and a final state, labeled by $i$ and $f$, can be written in terms of the corresponding WDFs as

$$i\hbar P_{if} = \langle W_f \mid L_{\text{int}} \mid W_i \rangle, \tag{7.3}$$

where $L_{\text{int}} = H_{\text{int}}\left(q + \dfrac{i\hbar}{2}\dfrac{\partial}{\partial p}, p - \dfrac{i\hbar}{2}\dfrac{\partial}{\partial q}\right) - H_{\text{int}}^{*}\left(q + \dfrac{i\hbar}{2}\dfrac{\partial}{\partial p}, p - \dfrac{i\hbar}{2}\dfrac{\partial}{\partial q}\right).$

The behavior of transition matrix elements in mixed quantum systems with a few degrees of freedom, with one or several regular islands in the ergodic sea, was studied by Boosé and Main [1996]. They showed that the mean associated to the distribution of diagonal transition matrix elements is a weighted sum of classical means over the ergodic part of PS and over the stable periodic orbit. The variance characterizing distributions of non-diagonal transition matrix elements can be well approximated with the weighted sum of Fourier transforms of averaged classical autocorrelation functions along an ergodic trajectory and along the stable periodic orbit.

In the formulation (7.1) of transition probabilities, the values of the two WDF over the common PS domain contribute. In some cases it is possible to identify a smaller PS domain that has a dominant contribution to the integral in (7.1). This is true especially when number states are the initial or final states of the transition, since the corresponding WDF, in the large-$n$ limit, has a prominent peak around the classical PS trajectory, a lower-amplitude oscillatory behavior inside it and an exponential decrease outside. Its WDF contribution in (7.1) is therefore expected to come mainly from the neighborhood of the PS trajectory. This is indeed the case when photon distributions for arbitrary states are calculated, or when probability amplitudes of transitions (due to a sudden change in conditions) between number states are considered. In this case the PS representations of the energy states as given in Fig.1a and the other involved state offer an intuitive explanation of the behavior of different quantities. Due to normalization reasons, the area of intersection between the two PS representations multiplied by $1/(2p\hbar)$ gives a good approximation to (7.1). If the PS overlap between the quantum states consists of more than one region, interference effects appear, as extensively discussed in Schleich [2001]. For two or more areas of overlap, the transition probability is obtained by adding the complex probability amplitudes associated to each area. The total transition probability between two states $|\boldsymbol{y}\rangle$ and $|\boldsymbol{c}\rangle$ is then given by

$$\langle \boldsymbol{c} | \boldsymbol{y} \rangle = \sum_j \left( \frac{j\text{th area of overlap}}{2p\hbar} \right)^{1/2} \exp\left[ \frac{i}{\hbar} \left( j\text{th area enclosed by central lines} \right) \right] \qquad (7.4)$$

With this area-of-overlap principle we can easily view the Poissonian energy spread of a coherent state for example, $P_{n,coh} = |\langle n | \boldsymbol{y}_{coh} \rangle|^2 = (\boldsymbol{a}^{2n}/n!)\exp(-\boldsymbol{a}^2)$ as arising from the overlap between the PS representations of the $n$-th eigenstate of the harmonic oscillator and of the wavefunction of the coherent state (Fig.6a). The main contribution to $\langle n | \boldsymbol{y}_{coh} \rangle$ arises from regions around the turning points $q_n = \sqrt{2(n+1/2)/\boldsymbol{k}}$, and so $\langle n | \boldsymbol{y}_{coh} \rangle$ follows in its $n$ dependence the wavefunction of the coherent state in the $q$ variable. When two coherent states of identical mean photon number but different phases are in a quantum superposition, squeezing, as well as sub-poissonian and oscillatory photon statistics can appear (Schleich, Pernigo and Kien [1991]).

This PS interpretation intuitively supports the mathematical result that the energy distribution for squeezed states gets narrower than the Poissonian (becomes sub-Poissonian) when we increase $s$ and that for strong squeezing $s \to \infty$ the distribution starts to oscillate with period two, since $P_{2n+1} = 0$, $P_{2n} \neq 0$ (Schleich, Walls and Wheeler [1988], Schleich [2001]). For the overlap between a number state and a displaced strongly-squeezed state (Fig.6b) with $\text{Im}\,\boldsymbol{a} = 0$, there are two areas of overlap with reduced areas of intersection $A_n/2p\hbar$, disposed symmetrically above and below the coordinate axis, and (7.4) reduces to $P_n = |(A_n/2p\hbar)^{1/2}\exp(i\boldsymbol{j}_n) + (A_n/2p\hbar)^{1/2}\exp(-i\boldsymbol{j}_n)|^2 = 4(A_n/2p\hbar)\cos^2\boldsymbol{j}_n$, where $\boldsymbol{j}_n = S_n - \boldsymbol{p}/4$, with $S_n$ the PS area enclosed by the vertical line at $\sqrt{2}\boldsymbol{a}$ and the Kramers center lines of the energy band $n$. In classical mechanics the probability amplitudes would be given by the reduced area of overlap, no phase factors being included. The two interfering quantum probability amplitudes in PS can be viewed as the analogous of the probability amplitudes in configuration space in the Young's double-slit experiment. The phase difference is now given by the PS area caught between the interfering states, whereas in Young's experiment it is determined by the difference in optical path length from the two slits to the point of detection. In the same limit of strong squeezing, $A_n$ $= \exp[-2(n+1/2 - a^2)/s]/[2p\pi(n+1/2 - a^2)]^{-1/2}$. The PS overlap principle also accounts for the behavior of the energy distribution of a rotated squeezed state, which has a rapid oscillation and a slow modulation on top of it, in contrast to the nonrotated case where there is a single but large period.

Examples of cases when more than two areas of overlap occur are the calculation of the photon number distribution of the squeezed number and squeezed thermal states (Kim, de Oliveira and Knight [1989]). In these cases highly structured number distributions are obtained due to interference effects from the four PS areas (see Fig.6c). In particular, if squeezed photon number states overlap photon number states with a different parity the interference is destructive and the photon number distribution vanishes. For overlaps between squeezed photon number states and photon number states with the same parity the interference is constructive and the photon number distribution has nonzero values. The area of overlap principle explains also the fact that the phase probability distribution of highly squeezed states undergoes a transition from a single- to a double-peaked shape when the product of squeeze and displacement parameters is decreased (Schleich, Horowicz and Varro [1989]).

Finally, it should be noted that when the PS overlap approach is extended to the hyperbolic space, the PS overlap areas should be replaced by weighted areas, since for the hyperbolic space the PS is not represented by the same embedded sheet as the configuration space (Chaturvedi, Milburn and Zhang [1998]). The hyperbolic space can be employed to characterize active interferometers, while passive interferometers can be described in a spherical space, for which the PS has the same spherical geometry.

## 8. CLASSICAL AND QUANTUM INTERFERENCE IN PHASE SPACE

Classical and quantum interferences are caused by the linear superposition principle of fields and quantum wavefunctions, respectively, which follow from the linearity of the corresponding wave equations. Young's type experiments, or one-photon interference experiments, can be successfully explained by either classical or quantum theory. Differences between the two theories can only be observed in experiments that involve the interference of intensities. In intensity interferometry experiments, or two-photon interferometry, as that of Hanbury-Brown and Twiss, the interference/correlations between the intensities of two electric fields detected by separate photomultipliers are measured. Classical theory predicts in this case an interference of intensities, whereas quantum theory treats the interference still at the level of probability amplitudes. The predictions of the two theories are thus different. The one-photon and two-photon interference experiments have been recently reviewed by Mandel [1999]. Interference experiments have been observed also for charged and neutral particles, spins, Bose-Einstein condensates, fluxons propagating in Josephson rings, atoms, experiments being even designed to demonstrate the nonlocal nature of the quantum interference. The theoretical and experimental work in the quantum interference domain is immense; to not do injustice by inevitably omitting valuable papers, we specifically refer only to those works directly related to the PS approach.

The correlation function between the electromagnetic field $\hat{E}(q,t) = i\sum_k (\hbar w_k/2e_0)^{1/2}[\hat{a}_k u_k(q)\exp(-iw_k t) - \hat{a}_k^+ u_k^*(q)\exp(-iw_k t)] = \hat{E}^+(q,t) + \hat{E}^-(q,t)$ at the space-time point $x = (q,t)$ and the field at $x' = (q',t')$ is defined as $G^{(1)}(x,x')$ $= Tr[\hat{r}\hat{E}^-(x)\hat{E}^+(x')]$. This first-order correlation function is sufficient to account for classical or quantum one-photon interference experiments. Ideal detectors working on an absorption mechanism yield a signal $I(q,t) = Tr[\hat{\imath}\hat{E}^-(q,t)\hat{E}^+(q,t)]$ (Walls and Milburn [1994]). Higher-order correlation functions are necessary to describe experiments involving intensity correlations. The $n$th order correlation function is defined as

$$G^{(n)}(x_1...x_n, x_{n+1}...x_{2n}) = Tr[\hat{\imath}\hat{E}^-(x_1)...\hat{E}^-(x_n)\hat{E}^+(x_{n+1})...\hat{E}^+(x_{2n})], \qquad (8.1)$$

whereas the $n$-fold delayed coincidence rate is proportional to $G^{(n)}(x_1...x_n, x_n...x_1)$. The $n$-th order correlation function satisfy the following properties:

$$G^{(n)}(x_1...x_n, x_n...x_1) \geq 0, \qquad (8.2a)$$

$$G^{(n)}(x_1...x_n, x_n...x_1)G^{(n)}(x_{n+1}...x_{2n}, x_{2n}...x_{n+1}) \geq |G^{(n)}(x_1...x_n, x_{n+1}...x_{2n})|^2. \qquad (8.2b)$$

The odd-ordered correlation functions contain information about the phase fluctuations of the electromagnetic field, no such information being contained in the even ordered correlation functions. The latter, including the second-order correlation function, are a measure of the fluctuations in the photon number. For fields propagating in nondispersive media no difference is made between longitudinal and temporal coherence. However, for dispersive propagation, as is the case for electrons or neutrons in vacuum, or for light propagating in a medium, one should distinguish between spatial and temporal coherence. A discussion on these two types of coherence in dispersive media can be found in Hamilton, Klein and Opat [1983].

## 8.1. CLASSICAL AND QUANTUM ONE-PHOTON INTERFERENCE

In one-photon Young's type interference experiments the intensity observed on the screen is given by $I = G^{(1)}(q_1,q_1) + G^{(1)}(q_2,q_2) + 2\operatorname{Re}[G^{(1)}(q_1,q_2)]$, where $q_1$, $q_2$ are the positions of the two pinholes. The first two terms describe the intensities from the individual pinholes, the interference fringes originating from the last, interference term, which can be rewritten as $2|G^{(1)}(q_1,q_2)|\cos \mathcal{J}(q_1,q_2)$. Note that

$$G^{(1)}(q,q') = \int dp \exp[-ip(q-q')/\hbar]W(p,(q+q')/2). \qquad (8.3)$$

The normalized first-order coherence function, defined as

$$g^{(1)}(q_1,q_2) = G^{(1)}(q_1,q_2)/[G^{(1)}(q_1,q_1)G^{(1)}(q_2,q_2)]^{1/2} \qquad (8.4)$$

is associated with the visibility of the interference fringes, given by $V = (I_{max} - I_{min})/(I_{max} + I_{min})$. More precisely, for incoherent fields for which $g^{(1)}(q_1,q_2) = 0$ no interference fringes appear, whereas full coherence, corresponding to $|g^{(1)}(q_1,q_2)| = 1$ is associated also with maximum fringe visibility. When the fields on each pinhole have equal intensities, $V = |g^{(1)}|$; the first-order coherence function was also identified with the degree of path indistinguishability (Mandel [1991]).

More generally, the fields for which the first- (higher-) order correlation function factorizes are first- (higher-) order coherent. Coherent states satisfy this criterion. In particular, the wavefunction of a coherent field incident on the pinholes 1 and 2 factorizes as $|a_1, a_2\rangle = |a_1\rangle|a_2\rangle$ and can therefore represent two independent light beams. Interference between independent light beams can occur if the phase relation between them varies slowly, and was observed for single-mode independent lasers by Pfleegor and Mandel [1967] even for light intensities so low that one photon is absorbed before the next is emitted by one or the other source. To have interference between independent light beams, it is necessary that the correlation function between the states of the radiation field does not vanish; this condition is not satisfied for example by the Fock states.

In terms of the WDF, the 'eventuality' of interference is described by the interference term, which appears even when the two interfering beams do not yet overlap. Wolf and Rivera [1997] showed that interference terms in the WDF, called 'smile function', exist also for superpositions of coherent classical optical fields, termed by an extension of language optical Schrödinger-cat states. The marginal projection of the smile yields the transmissivity of the physical hologram obtained by superposing the two beams. Unlike classical optical experiments, where the interference pattern appears immediately, when few photons pass through the set-up at a certain time, the quantum interference pattern reveals itself in time. The fringe visibility is however the same as with high-intensity light sources (see for example Franson and Potocki [1988]).

Referring to a two-slit experiment, in which the coherent fields immediately after the slits have a Gaussian form, the interference term in the WDF is present even immediately behind the slits, when the coherent beams do not overlap in real space (see Fig.7a and also (5.3) for the WDF of a superposition of two coherent states). Interference in the $q$ or $p$ domain arise when the outer terms in the WDF, representing the interfering fields, have a common projection interval along the respective axis. So, immediately after the slits the beams interfere only along the $p$ axis, interference in real space occurring after propagation through a sufficiently long distance in free space, such that the outer terms in the WDF begin to have a common projection interval over the $q$ axis also. At propagation through free space the WDF suffers a shear transform, as can be seen in Fig.7b, which does not affect the interference pattern along the $p$ axis. Experimental demonstration of $p$-space interference, in the absence of $q$-space interference, as well as the conclusion that interference should be treated rather in PS than in the configuration space, has been already provided by Rauch [1993] and Jacobson, Werner and Rauch [1994]. The distinction between interaction and interference is best described in PS: there is interaction (transition) if the WDFs of two states overlap, and there is interference if the corresponding WDFs have common projections along $q$ or $p$, but are still well separated.

Of course, interference patterns can only be observed when no knowledge about the slit the quantum particles or classical waves go through is available. In the quantum case the interference fringe visibility $V$ and which-way knowledge $K$ can be related through $V^2 + K^2 \leq 1$ (Schwindt, Kwiat and Englert [1999]). Experimental results for pure, mixed and partially-mixed input states show good agreement with the theoretical prediction $V^2 + K^2 = 2Tr(\hat{r}^2) - 1$. In PS the measurement process, including the which-path measurements, can be viewed as PS filtering, the WDF of the incoming light being filtered not only by the measurement devices, but by any part of the set-up in which they are propagating (Dragoman [2001a]). In particular, when one slit is closed, or when a which-path measurement is performed, the PS area occupied by the system is reduced to only one region (corresponding to the remaining beam), which has no other domain to interfere with. Interference is thus lost by PS filtering. The central role of the interference term in the WDF in quantum interference was also evidenced by Wallis, Röhrl, Naraschewski and Schenzle [1997]. They studied the macroscopic interference of two independent Bose-Einstein condensates and their PS dynamics, showing that the distance of interference fringes varies linearly in time with a velocity inversely proportional to the distance between the two condensates and that collisions reduce the fringe visibility.

The destruction by a which-path measurement of the interference fringes in a double-slit interference, with $d$ the slit separation, can also be interpreted as a disturbance of the particle's momentum by an amount of at least $p\hbar/2d$, in accordance with the uncertainty principle. Wiseman, Harrison, Collett, Tan, Walls and Killip [1997] showed that this momentum transfer caused by the interaction with the which-path measuring apparatus need not be local; i.e. need not act at either of the slits through which the particle passes. In a local interaction, as in Einstein's recoiling slit or Feynman's light microscope, the momentum transfer is modeled by random classical momentum kicks which have the same effect on the momentum distribution of a particle passing through a single slit, and so the single-slit diffraction pattern is also smeared, in the same way as the double-slit interference pattern. In a nonlocal momentum transfer, as in the schemes proposed by Scully, Englert and Walther [1991] or Storey, Collett and Walls [1993], the single-slit diffraction pattern is not broaden, and, unlike classical momentum kicks, the quantum momentum transfer depends in general on the initial wavefunction of the particle. The PS representation of both cases can be readily given in terms of the WDF of the (classical or quantum) momentum-transferring device, $W_t(q,p)$, which changes the initial WDF $W_i(q,p)$ into a final one $W_f(q,p)=\int dp'W_i(q,p-p')W_t(q,p')$. (A 'momentum filtering' rather than a 'PS filtering' is considered.) The smearing of the diffraction pattern and the destruction of the interference fringes are caused by different momentum transfers, a local $P_{loc}(p)$ and a nonlocal one $P_{nonloc}(p)$; only $P_{nonloc}(p)$ cannot be less than $p\hbar/2d$. In a double-slit experiment, with very narrow slits separated by $d$, the interference term in the initial WDF is ruined by a nonlocal momentum transfer with a pseudoprobability distribution $P_{nonloc}(p)=W_i(0,p)$ (at $q$ midway between the slits, the particle is never found!). The visibility of the interference fringes is then changed to $\vee=\int dpP_{nonloc}(p)\exp(ipd/\hbar)$ ($V=|V|$ is the usual visibility). A which-path measurement device can change the visibility from 1 (the value in the absence of any which path measurement) to any less than unity value, up to zero for a perfect which-path measurement. In a classical momentum-transfer experiment $W_t(q,p)$ is the probability distribution for a particle at position $q$ to receive a momentum transfer $p$, so that the local momentum transfer when the particle is localized at the two slits is given by $P_{loc}(p)=(1/2)[W_t(d/2,p)+W_t(-d/2,p)]$. It is not related to the WDF interference term and so plays no role in the destruction of interference, but determines the diffraction pattern of a particle that is in a classical mixture of being at the two slits.

This interpretation of the destruction of the interference pattern by the nonlocal momentum transfer is indirectly supported by the Aharonov-Bohm effect. In this quantum, nonlocal effect the interference fringes move with the applied field within the constant diffraction envelope. A local, classical momentum transfer would shift the entire pattern, similar to the way a local which-path measurement smears the interference pattern. A direct interpretation of the Aharonov-Bohm effect in terms of the WDF is given in Dragoman [2001b]. If the two interfering beams of charged particles acquire different phases due to a non-vanishing vector potential, the interference term in the WDF shifts also, while the individual WDF terms remain the same (see Fig.7c). According to the interpretation in Dragoman [2001a], the interference pattern (given by the common projection intervals along the $q$ or $p$ domains of the WDFs of the interfering beams) remains the same, while the interference fringes, defined by the interference term in the WDF, move with the vector potential. It is worth mentioning that, at least in this case, the PS interpretation has offered a new prediction: there is an Aharonov-Bohm term (a shift of the interference fringes inside the same interference pattern) in the $p$-space also, not only in the $q$-space. This prediction follows readily from Fig.7c.

Quantum and classical interference is mostly studied using the WDF. Other PS quasidistribution functions can be used to this end. Chountasis and Vourdas [1998] advocated the advantages of the Weyl function for the study of interference. The Weyl function, defined as $\tilde{W}(Q,P) = Tr[\hat{r}\hat{D}(Q,P)]$, where $\hat{D}(q,p)$ is the displacement operator, is related to the WDF through a double Fourier transform: $\tilde{W}(Q,P) = \iint dqdpW(q,p)\exp[-i(Pq-pQ)]$. This can be seen easier writing the WDF as $W(q,p) = \boldsymbol{p}^{-1}Tr[\hat{r}\hat{U}(q,p)]$, where $\hat{U}(q,p) = \hat{D}(q,p)\hat{U}_0\hat{D}^+(q,p)$ is the displaced parity operator, with $\hat{U}_0 = \exp(i\boldsymbol{p}a^+a)$. The Weyl function is a generalized correlation function, the latter being a special case of the Weyl function for $P = 0$ or $X = 0$. Whereas for a superposition of states labeled by $i$ the WDF auto-terms are located around the points $(\langle q_i \rangle, \langle p_i \rangle)$, they are located around the origin for the Weyl function. For a Schrödinger cat state, for example, the two Gaussian auto-terms in the WDF, and the oscillating cross-term term in the middle are mirrored into an oscillating term of the Weyl function around the origin representing auto-terms and two Gaussians representing cross-terms. Note that the correspondent in classical optics of the Weyl function is the ambiguity function. A more sophisticated engineering of the PS position of the auto and interference terms can be achieved using fractional Fourier operators. The fractional Fourier operator $\hat{V}(\boldsymbol{q}) = \exp(i\boldsymbol{q}\hat{a}^+\hat{a})$ rotates the momentum and position operators in phase space with an angle $\boldsymbol{q}$: $\hat{q}_q = \hat{V}(\boldsymbol{q})\hat{q}\hat{V}^+(\boldsymbol{q}) = \hat{q}\cos\boldsymbol{q} + \hat{p}\sin\boldsymbol{q}$, $\hat{p}_q = \hat{V}(\boldsymbol{q})\hat{p}\hat{V}^+(\boldsymbol{q}) = -\hat{q}\sin\boldsymbol{q} + \hat{p}\cos\boldsymbol{q}$, and can be used to define generalized WDFs $W(q,p;\boldsymbol{q}) = Tr[\hat{r}\hat{U}(q,p;\boldsymbol{q})]$, where $\hat{U}(q,p;\boldsymbol{q}) = \hat{D}(2q,2p)\hat{V}(\boldsymbol{q})$. The WDF and Weyl functions are particular cases for $\boldsymbol{q} = \boldsymbol{p}$ (for which $\hat{V}(\boldsymbol{p}) = \sum_{n=0}^{\infty}(-1)^n|n\rangle\langle n|$ is the parity operator) and $\boldsymbol{q} = 0$, respectively. For a Schrödinger cat state, as $\boldsymbol{q}$ decreases from $\boldsymbol{p}$ to 0 the auto-terms in the generalized WDF change their form from Gaussians to oscillatory and move in the PS plane ending at the origin, whereas the cross-terms, which are originally oscillatory, become Gaussians. The angle $\boldsymbol{q}$ can even be complex, in which case its positive imaginary part describes attenuation ('filtering'), while its negative imaginary part represents amplification (Chountasis, Vourdas and Bendjaballah [1999]).

Multi-dimensional interference patterns in the quantum probability or classical field intensity distribution, called intermode traces (or quantum carpets), can appear due to pair interference between individual eigenmodes of the system (Kaplan, Stifter, van Leeuwen, Lamb jr. and Schleich [1998]). The resulting interference pattern, strongly pronounced if the intermode traces are multi-degenerate, can be observed in many areas of wave physics, as for example for confined quantum particles, atoms scattered at a periodic laser-induced grating, electromagnetic waveguides or light diffraction. The similitude of the quantum and classical interference pattern is based on the mathematical analogy between the Schrödinger equation for the wavefunction of a quantum particle and the Maxwell's equations of classical electrodynamics under the paraxial, fixed polarization (scalar theory) approximation. This quantum-classical similarity can be exemplified by the fact that an electron in a quantum box with infinite walls is analogous to an electromagnetic wave in a waveguide with metallic walls, or by the fact that electrons moving in periodic potentials (in the conduction band of solids, for example) behave as almost free particles. In these two cases the similarity between the classical and the quantum case can be evidenced also using the WDF (see Marzoli, Friesch and Schleich [1998] and Lee [1995a], respectively).

## 8.2. CLASSICAL AND QUANTUM TWO-PHOTON INTERFERENCE

Intensity correlation experiments of the Hanbury-Brown and Twiss type measure the joint photocount probability of detecting the arrival of a photon at time $t$ and another at time $t + \boldsymbol{t}$, given by

$$G^{(2)}(\boldsymbol{t}) = \langle \hat{E}^-(t)\hat{E}^-(t+\boldsymbol{t})\hat{E}^+(t+\boldsymbol{t})\hat{E}^+(t)\rangle. \tag{8.5}$$

The two-time photon-number correlations can be measured on a sub-picosecond scale using dual-pulse, phase-averaged, balanced homodyne (McAlister and Raymer [1997]). Even the cross-correlations and mutual coherence of optical and matter fields can in principle be measured (Prataviera, Goldstein and Meystre [1999]).

Second-order coherence can be also characterized by the normalized second-order correlation function $g^{(2)}(\boldsymbol{t}) = G^{(2)}(\boldsymbol{t})/|G^{(1)}(0)|^2$. For fields which posses second-order coherence, as for laser fields, $G^{(2)}$ factorizes and $g^{(2)} = 1$, independent of the delay. Coherent fields exhibits Poissonian statistics. The correlation function factorizes always for $\boldsymbol{t} \gg \boldsymbol{t}_c$ (correlation time of light), whereas for shorter times $g^{(2)}$ can be higher or lower than the value for coherent light. In the first case, there is a high probability that a second photon will be detected arbitrarily soon, phenomenon known as photon bunching. This phenomenon occurs also for classical electromagnetic fields with fluctuating amplitudes for modes; in particular, for thermal light $g^{(2)}(0) = 2$ and approaches unity from above as a function of $\boldsymbol{t}$. For the thermal light the photon distribution is super-Poissonian. However, the fields for which $g^{(2)}(0) < 1$ (for example Fock states, or the light emitted by an atom driven by a laser field, for which $g^{(2)}(0) = 0$) are regarded as quantum; the corresponding photon antibunching phenomenon cannot be predicted by a classical theory since negative probabilities would be required. For quantum fields $g^{(2)}(0) = 1 + (V(n) - \overline{n})/\overline{n}^2$, with $V(n) = \langle(\hat{a}^+\hat{a})^2\rangle - \langle\hat{a}^+\hat{a}\rangle^2$. Antibunching and squeezing are exclusive quantum properties; they are not exhibited by fields with a positive $P$ distribution. The coherent states, characterized by a delta-like $P$ distribution, define the boundary between classical and quantum behavior. The fields for which $g^{(2)}(\boldsymbol{t}) < 1$ are characterized by a sub-Poissonian statistics; however, it is possible to have fields that can exhibit super-Poissonian statistics over some time interval, but for which $g^{(2)}(\boldsymbol{t}) > g^{(2)}(0)$.

A review of the single and double beam experiments, which measure the degree of second-order coherence, and thus can discriminate classical against quantum theories can be found in Reid and Walls [1986]. Classical models predict for two-photon interference a maximum of 50% visibility, as does quantum theory for experiments that involve only the wave nature of radiation. However, visibilities greater than 50% are predicted by quantum theory for certain nonlocal entangled states, called Einstein-Podolsky-Rosen (EPR) states. For these states, with no classical analog, the interference visibility can ideally reach 100%. The experimental confirmation of nonclassical values for two-photon interference visibility has been provided, for example, by Shih, Sergienko and Rubin [1993]. Two-photon interference for entangled photons, obtained from type II down-conversion, have been observed in experiments in which the photons arrive at the beam splitter at much different times. This confirms the fact that for entangled photons the interference pattern cannot be viewed as produced by the interference between two individual photons (Pittman, Strekalov, Migdall, Rubin, Sergienko and Shih [1996]). Two-photon interference effects were even observed when the entangled photon pairs were generated from two laser pulses well separated in time. This effect, not expected classically, shows a visibility that depends on the delay time, reaching a maximum value of 50%. However, visibilities up to 100% can be obtained for multiple pulses delayed in time with respect to each other (Kim, Chekhova, Kulik and Shih [1999]). The applications of the WDF for the study of two-mode quantum correlations, in particular the cross-correlations between the two modes that violate classical inequalities, are discussed in Walls and Milburn [1994].

In particular, Bell's theorem (Bell [1991]) provides a test of the predictions of the whole class of local hidden variable theories against quantum mechanics. Bell's inequalities and the EPR paradox demonstrate the nonlocality of quantum mechanics as expressed by the correlations between different subsystems of an entangled quantum system, for which the eigenstate does not factorize in the eigenstates of the subsystems. Bell correlations in PS can be tested with quantum optical means (Leonhardt and Vaccaro [1995]). Banaszek and Wódkiewicz [1999] showed that the correlation functions that violate Bell's inequalities for correlated two-mode quantum states of light, are equal to the joint two-mode $Q$ function and the WDF. The connection between the nonlocality of entangled states and the $Q$ function is surprising since the positiveness of the $Q$ function is usually considered as a loss of quantum properties due to simultaneous measurement of canonically conjugated observables (see section). The non-positivity of the WDF, on the other hand, cannot be unequivocally related to violations of local realism, since in some cases Bell's inequalities are violated by states with a positive WDF (Banaszek and Wódkiewicz [1998]). So, the positivity or non-positivity of the WDF cannot act as criterion for the locality or nonlocality of quantum correlations. Moreover, using the WDF, the EPR correlated state of spin 1/2 systems can be treated analogously to a local hidden variable model if the probability distribution function is allowed to become negative (Agarwal, Home and Schleich [1992]). Negative conditional probabilities between nonorthogonal polarization components of entangled photon pairs are responsible for the violation of Bell's inequalities. These negative probabilities can be observed in finite-resolution measurements of the nonclassical polarization statistics of entangled photon pairs, or in finite-resolution measurements of the polarization of a single photon (Hofmann [2001]).

Entangled photon states (in time, frequency, direction of propagation, or polarization) can be created in nonlinear optical processes such as parametric downconversion. The second-order correlation functions of two-photon wave packets entangled in polarization and space-time can be studied using the WDF formalism. For example, in quantum-beating experiments (Ben-Aryeh, Shih and Rubin [1999]) the coincidence-counting rate is proportional to the integral of the relative two-photon wave probability distribution (in relative coordinates) over the retarded time difference, or to an integral over the frequencies difference. Both momentum-frequency and position-time coordinates can be accounted for in the WDF picture, the WDF for the two-particle entangled states being a multiplication of a WDF that depends on the relative coordinates of the two-photon with a WDF that depends on the central coordinates of the two-photon.

Entangled two-photon pairs, or biphotons, generated by spontaneous parametric down-conversion display some properties that are similar to those of photons generated by incoherent sources. For example, the two-particle wavefunction and the spatial pump-field distribution in the biphoton case are analogous to the second-order coherence function and the source intensity distribution in the incoherent case. Moreover, the van Cittert-Zernike theorem, valid for incoherent optical sources, and the partial-coherence theory of image formation have counterparts for biphotons. If we compare, however, the photon count rate in the incoherent case with the biphoton count rate in the entangled-photon case, a duality rather than analogy is observed, similar to the duality between single and two-photon interference of biphotons (Saleh, Abouraddy, Sergienko and Teich [2000]). This duality originates in the fact that the separability of the second-order coherence function is associated with high-visibility ordinary interference fringes, whereas separability in the biphoton wavefunction is associated with the absence of entanglement and so low-visibility biphoton interference fringes. The biphoton interference fringe visibility $V_{12}$ is related to the visibility $V_{inc}$ of single-photon fringes created by an equivalent ordinary incoherent source of the same source distribution by $V_{12} = (1 - V_{inc}^2)/(1 + V_{inc}^2)$. It can also be related to the visibility $V_1$ of the interference pattern due to signal (or idler) photons (marginal single-photon patterns) by $V_1^2 + V_{12}^2 = 1$.

## 9. UNIVERSALITY OF THE PHASE SPACE TREATMENT

In the previous sections we have advocated the advantages of a PS approach to the problem of finding the correspondence between classical optics and quantum mechanics. It was shown that classical wave optics can be treated in PS via quasidistribution functions with similar properties to the quantum quasidistribution functions. Ray optics becomes then the 'classical' limit of wave optics in the same sense as classical mechanics is the limit of quantum mechanics. In this last section we would like to emphasize the universality of the PS treatment. It is not only employed for the study of classical mechanics, classical optics or quantum mechanical systems, but also for the study of statistical mechanics and particle optics, for example. The significance of the classical-quantum correspondence broadens considerably by realizing the universal character of the PS approach.

For example, the PS concepts can be employed for the study of the propagation of electronic rays. In this case a WDF can be defined as in quantum mechanics, with the beam emittance playing the role of Planck's constant and the longitudinal coordinate of propagation playing the role of time. The electronic beams can be transported, analogous to optical rays, with combinations of quadrupoles, drift sections, bending magnets, etc. The first two devices are the electronic counterparts of optical lenses and free space sections, respectively. More complicated optical set-ups, which can realize for example squeezing along a tilted direction in PS, can be implemented in electronic optics. In particular, a set-up for measuring the Wigner angle of rotation for electron beams has been proposed by Ciocci, Dattoli, Mari and Torre [1992].

A Wigner PS representation for a reduced density operator can also be introduced in thermo-field-dynamics for the study of thermal excitation (Berman [1990]). The thermal excitation can then be viewed as a temperature-dependent radial spreading in PS.

Uncertainty relations of the form $\Delta E^{a} \Delta i_{b} \geq k_{B} d_{b}^{a}$ can also be encountered in statistical mechanics (Gilmore [1985]), as in any physical theory that can be formulated in terms of canonically conjugated variables. In this case the variance is defined as $\Delta x = \langle (x - \bar{x})^{2} \rangle^{1/2}$, where $x$ is the random variable with mean value $\bar{x} = \langle x \rangle$, $E^{a}$ is an extensive thermodynamic variable characterizing the system, and $i_{b} = \partial S / \partial E^{b}$ is the intensive thermodynamic variable conjugated to $E^{b}$ in the entropy representation. For example, $\Delta U \Delta (1/T) \geq k_{B}$, $\Delta V \Delta (P/T) \geq k_{B}$, $\Delta N \Delta (m/T) \geq k_{B}$. A similar relation $\Delta E^{a} \Delta i'_{b} \geq k_{B} T d_{b}^{a}$, with $i'_{b} = \partial U / \partial E^{b}$, holds in the energy representation, where $\Delta S \Delta T \geq k_{B} T$, $\Delta V \Delta P \geq k_{B} T$, $\Delta N \Delta m \geq k_{B} T$. The factor ½ is missing in these uncertainty relations compared to those in quantum mechani cs and signal processing, because the conjugacy relation are obtained here taking derivatives of a probability distribution function, not of probability amplitudes as in quantum mechanics. The uncertainty relations of statistical mechanics express the duality between probability and statistics, and are equivalent to the stability relations of equilibrium thermodynamics. As for quantum mechanics, where the classical limit corresponds to $\hbar \to 0$, the classical limit of statistical mechanics – thermodynamics – is obtained for $k_{B} \to 0$. In both cases the classical limits are characterized by the lack of the uncertainty relations.

The coherent and squeezed states, although defined and employed mostly in quantum optics, can be generated also in other domains. For example, the eigenstates of a generalized thermal annihilation operator, constructed using thermofield dynamics, are the fermionic coherent states, fermionic squeezed states, and their thermalized counterparts (Chaturvedi, Sandhya, Srinivasan and Simon [1990]). Coherent and squeezed states of phonons have also been investigated (Hu and Nori [1996]). These coherent phonon states can be excited phase coherently in Brillouin and Raman scattering experiments, or piezoelectric oscillators can generate coherent acoustic waves up to $10^{10}$ Hz. And the examples can continue.

Since PS methods involve the same mathematical language in all these different domains: wave optics, quantum optics, statistical mechanics, thermofield dynamics, etc., it is to be expected that our understanding of the origin of the formal relations between them, and our ability to constructively speculate these similarities will improve.

## 10. CONCLUSIONS

Throughout this paper we have revealed similarities and differences between quantum mechanics and classical optics. We have seen that the classical limit of the WDF, i.e. the classical probability distribution function, is not the same with the WDF in wave optics. The latter is an exact replica, from the point of view of definition, properties, and even mathematical expressions for certain fields, to the quantum WDF, if the Planck's constant is replaced by the wavelength of light. There is, however, a major difference between the quantum and optical WDF from the point of view of measurement. The quantum WDF cannot be directly measured with arbitrary accuracy, whereas the optical WDF can. Therefore, it is possible to produce classical waves with the same form as quantum wavefunctions (this can be done for any superposition of coherent or Fock states), generate their WDF by classical optical means, and study its propagation through classical optical set-ups that mimic quantum systems. As regards the behavior of the WDF, the results would be identical with those obtain in quantum mechanics, with the additional advantages of easier production and accurate measurement. This is interesting conceptually, because the starting point – classical fields and quantum states – have very different properties. In particular, the quantum interference principle holds for any superposition of quantum states, whereas in the classical case only overlapping fields interfere. Even the nature and significance of interference is different: for a quantum superposition of states the quantum particle is with a certain probability in only one of these states; one can measure, by repeating the experiment, only these probabilities. On the other hand, in a classical superposition the interference pattern appear as if the light is in both (or all, in general) slits at the same time, not in one or in the other. Field amplitudes are real objects in the classical world, but only probabilities in quantum mechanics. Why is it then, that the WDF washes away the difference in behavior and retains only the difference in the measurement procedure? The answer is in its bilinear character. Due to it, interference terms in the WDF appear even if the coherent classical beams do not really overlap; the interference in PS has the same character in both quantum and classical world. Quantum interference in PS can be mimicked by classical interference in PS, although the corresponding classical fields behave differently than their quantum counterparts.

This similitude holds for the (quantum and classical) wave theories, for which the PS can be considered as being partitioned in adjacent, interacting, finite-area cells, occupied by elementary Gaussian beams (with positive WDFs), such that whenever the occupied PS area exceeds the minimum value allowed by the uncertainty principle the PS interference between neighboring cells lead to negative values of the WDF. On the contrary, in the classical limit of quantum mechanics, or in geometrical optics the PS is continuous and a classical mechanical ensemble or light beam can be decomposed in a number of perfectly localized, non-interacting particles or rays. To get the illusion of classical mechanics the quantum uncertainties should be small on the observation scale (Royer [1991]).

A last, but not least important remark is that, due to the probabilistic character of the quantum wavefunction, which reflects the wave-particle duality, a quantum system can be meaningful compared to a classical one only when the particle or the wave character is involved, but not both. There are no classical states similar to the EPR entangled states, although the WDF can be employed even in this case in quantum mechanics. The quantum WDF cannot always be mimicked by a classical optical WDF, despite the evident similarities between them. The quantum WDF, and the quantum PS quasidistribution functions in general, are a combination of their classical optical counterparts, photon counting, the mysterious duality,.... When we would know all the ingredients and their proportions, quantum theory would not look so beautiful.

FIGURES

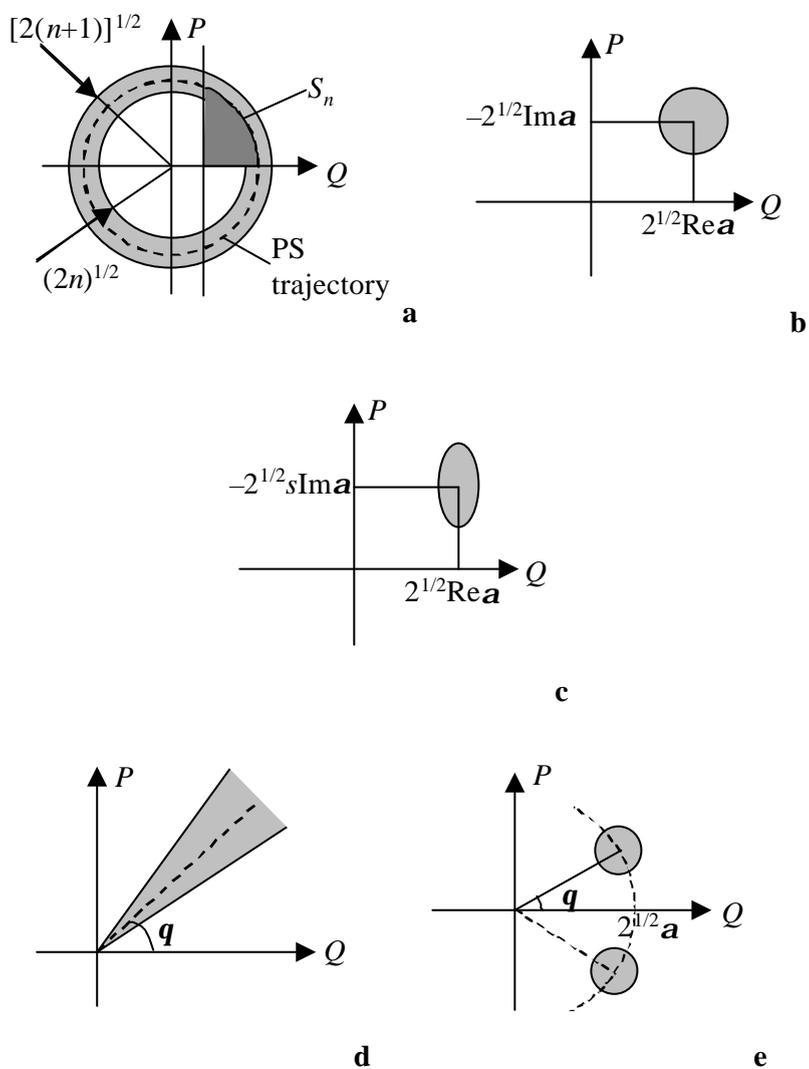

Fig.1    PS representations of (a) the $n$th energy eigenstate of a harmonic oscillator, (b) a coherent state, (c) a displaced squeezed state, (d) a phase state, and (e) a superposition of two coherent states

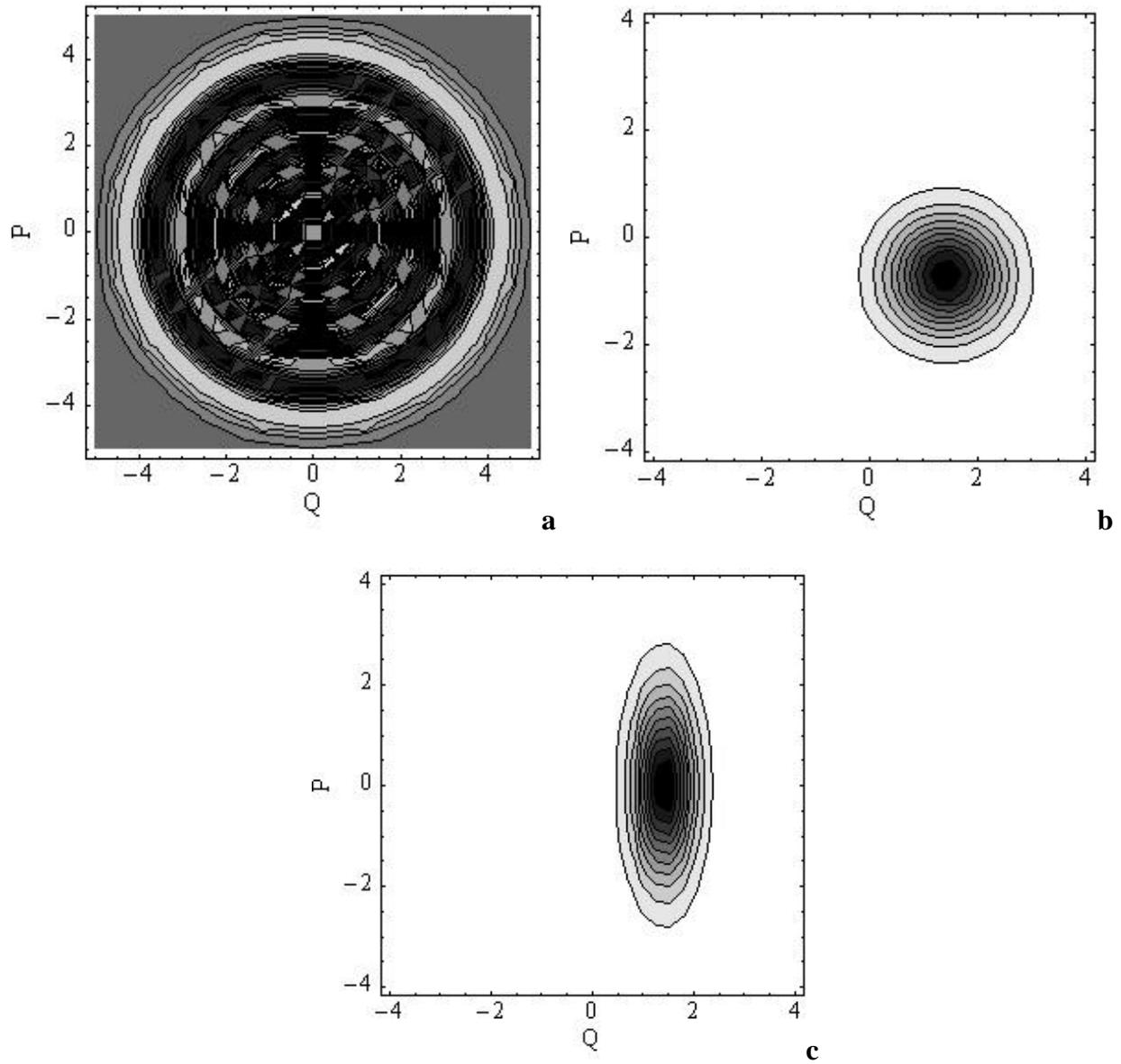

Fig.2    Contour plots of the WDF of (a) the $n = 10^{th}$ energy eigenstate of a harmonic oscillator, (b) a coherent state with Re $\boldsymbol{a}$ = 1, Im $\boldsymbol{a}$ = 0.5, (c) a squeezed state with Re $\boldsymbol{a}$ = 1, Im $\boldsymbol{a}$ = 0, $s$ = 3. The lighter areas correspond to higher values of the WDF in (a) and to lower values of the WDF in (b) and (c)

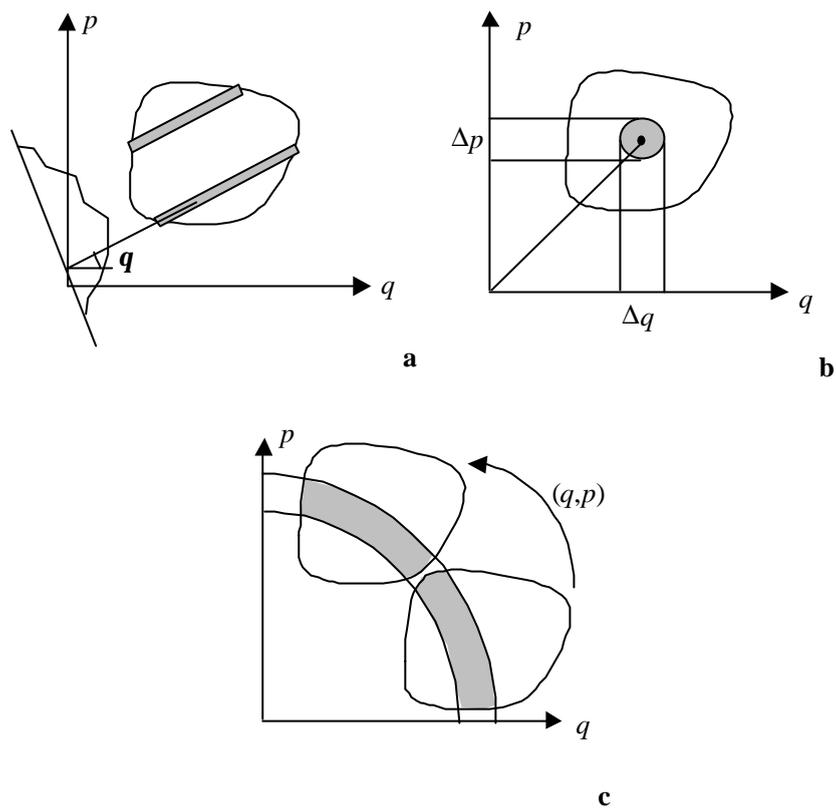

**a**

**b**

**c**

Fig.3    Schematic representation of the (a) tomographic method, (b) simultaneous measurement method, and (c) the ring method for the determination of the WDF

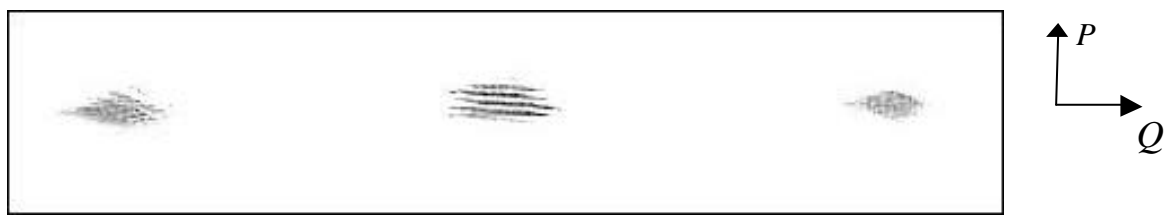

Fig.4    Negative image of the experimentally determined modulus of the WDF for a superposition of two coherent and spatially separated Gaussian beams

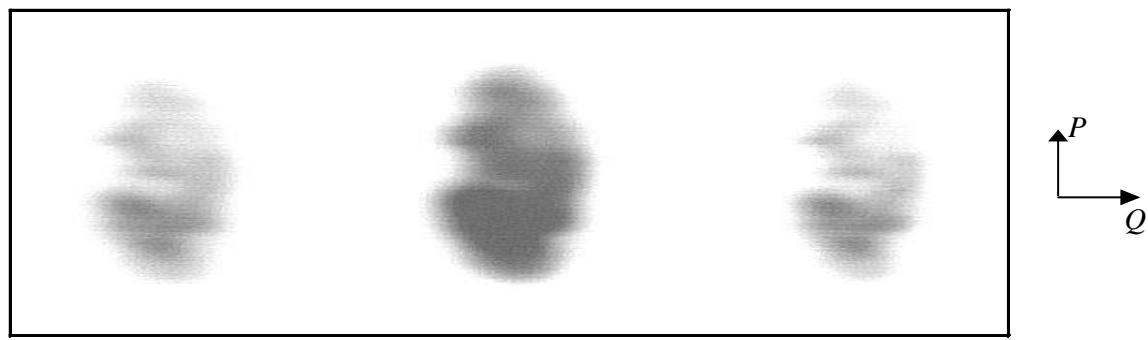

Fig.5     Negative image of the experimentally determined modulus of the WDF for a superposition of two incoherent and spatially separated beams

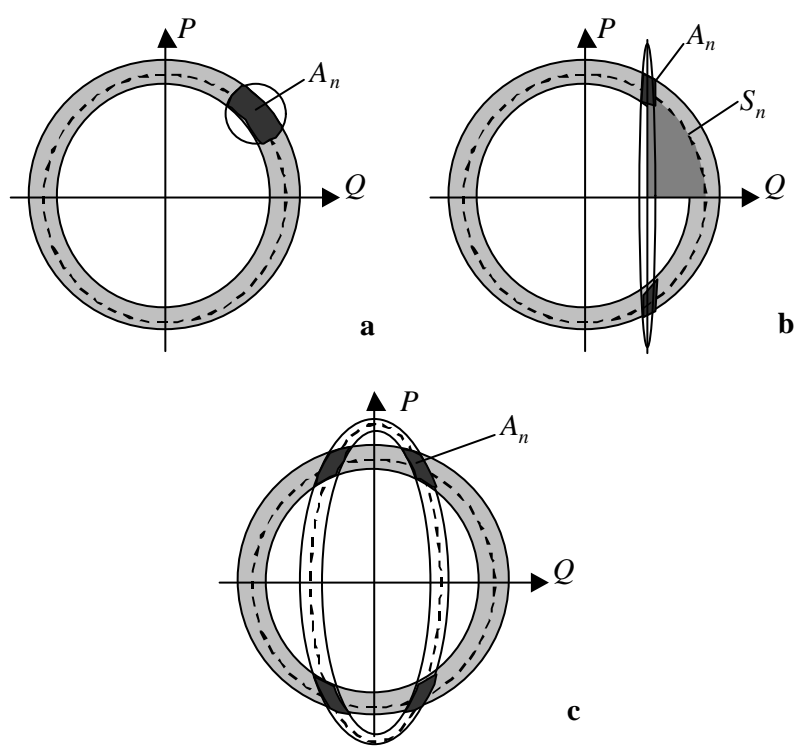

Fig.6     PS overlap between the $n$th energy eigenstate of a harmonic oscillator and (a) a coherent state, (b) a squeezed coherent state with real $\boldsymbol{a}$, and (c) a squeezed energy eigenstate

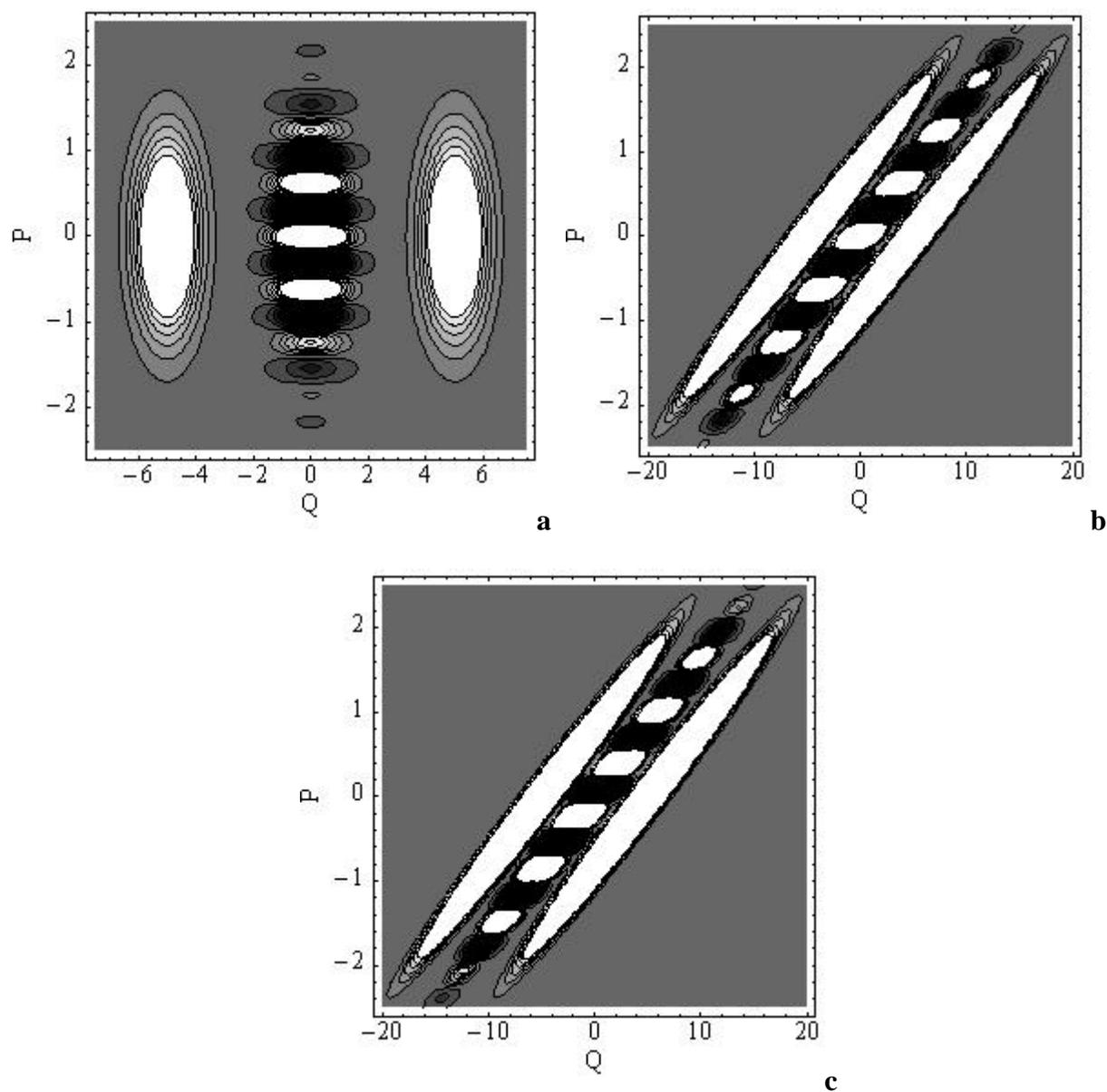

Fig.7      Contour plots of the WDF of a superposition of two Gaussian fields (a) immediately after the slits, and (b) after propagation through a sufficiently large distance for the beams to overlap in real space. (c) Same as (b) if the interfering beams acquire different phases. Lighter areas correspond to higher values of the WDF.